\documentclass[onecolumn,showpacs]{revtex4}

\usepackage{graphicx}
\newcommand{\gsim}{\mbox{{\raisebox{-0.4ex}{$\stackrel{>}{{\scriptstyle\sim}}
$}}}}
\newcommand{\lsim}{\mbox{{\raisebox{-0.4ex}{$\stackrel{<}{{\scriptstyle\sim}}
$}}}}

\begin{document}
\title{The spectral and polarization characteristics of the nonspherically decaying radiation
generated by polarization currents with superluminally rotating distribution patterns}
\author{H. Ardavan$^1$, A. Ardavan$^2$ and J. Singleton$^3$}
\affiliation{$^1$Institute of Astronomy, University of Cambridge,
Madingley Road, Cambridge CB3 0HA, United Kingdom\\
$^2$Clarendon Laboratory, Department of Physics, University of Oxford,
Parks Road, Oxford OX1 3PU, United Kingdom\\
$^3$National High Magnetic Field Laboratory, TA-35, MS-E536,
Los Alamos National Laboratory, Los Alamos, NM87545, USA}
\begin{abstract}
We present a theoretical study of the emission from a
superluminal polarization current whose distribution
pattern rotates (with an angular frequency $\omega$)
and oscillates (with a frequency $\Omega $) at the same time, and
which comprises both poloidal and toroidal components.
This type of polarization current is found in recent practical
machines designed to investigate superluminal emission.
We find that the superluminal motion of the distribution pattern of the emitting current generates
localized electromagnetic waves that do not decay
spherically, i.e.\ that do not have an intensity diminishing like ${R_P}^{-2}$
with the distance $R_P$ from their source.
The nonspherical decay of the focused wave packets that are
emitted by the polarization currents does not contravene conservation of energy:
the constructive interference of the constituent waves of such propagating
caustics takes place within different solid angles on spheres of different radii ($R_P$)
centred on the source.
For a polarization current whose longitudinal distribution (over an azimuthal interval of length $2\pi$) consists of $m$ cycles of a sinusoidal wave train, the nonspherically decaying part of the emitted radiation contains
the frequencies $\Omega \pm m\omega$; i.e.\
it contains {\it only} the frequencies
involved in the creation and implementation of the source.
This is in contrast to recent studies of the spherically decaying emission,
which was shown to contain much higher frequencies.
The polarization of the emitted radiation is found to be
linear for most configurations of the source.
\end{abstract}

\maketitle
\section{Introduction}
The electromagnetic field of a moving charged particle whose speed exceeds the speed of light
{\it in vacuo} is the subject of several papers written by Sommerfeld in 1904 and 1905~\cite{HA1},
papers which were regarded to have been superseded by special relativity soon after
their publication.  One reason for abandoning further investigations of the work initiated by
Sommerfeld, of course, was that any known particle that had a charge also had a rest
mass and so was barred from moving faster than light by the requirements of special relativity.
But there was an additional reason.  Not even a massless particle can move faster than light
{\it in vacuo} if it is charged; for if it does it would give rise to an infinitely strong
electromagnetic field on the envelope of the wave fronts that emanate from it.

It was not until the appearance of the works of Ginzburg and his coworkers~\cite{HA2,HA2A,HA2B}
that it was realized that, though no superluminal source of the electomagnetic field can
be point-like, there are no physical principles disallowing faster-than-light
sources that are {\it extended}.
The coordinated motion of aggregates of subluminally moving charged particles of opposite sign
can give rise to macroscopic polarization currents whose distribution patterns move superluminally.
The electromagnetic field that is generated by such extended sources,
on the other hand, may be built up by the superposition of the fields of their constituent
volume elements, elements which individually act as the superluminally moving point sources
considered by Sommerfeld~\cite{HA3,HA3A,HA3B,HA3C}.

The purpose of the present paper is to examine the radiation field of a particular class of such
volume-distributed sources:  polarization currents whose distribution patterns have the time dependence
of a travelling wave with a centripetally-accelerated superluminal motion.
A motivation for this work is the recent design, construction and testing
of experimental machines with this
characteristic~\cite{AA1,AA2,AA3,AA4,AA5,AA6}.

The Green's function for the problem we analyze is the familiar field of a uniformly rotating point source.
This is the Li\'enard-Wiechert field that is encountered in the analysis of synchrotron radiation,
except that here we do not restrict the speed of the source to the subluminal regime.
This uniformly rotating point source is used as a basic volume element of the extended superluminal source,
which is then treated in detail by superposing the Li\'enard-Wiechert fields of its constituent elements.
We find that fundamentally new radiation processes come into play as a result of lifting the
restriction to the subluminal regime,
processes that have no counterparts in either the synchrotron or the \v Cerenkov effects.

This paper is organized as follows. Section~II presents an introductory description of
the principles behind our calculation of the emission. It includes a definition of 
frequencies (Table~I) and polarization (Section IIE) of the source,
which are later used to discuss the spectral characteristics and polarization of the
emitted radiation.
A detailed mathematical treatment is given in Section III, with the
expression describing the nonspherically decaying component of the
radiation [Eq.\ (57)] being derived and discussed in Section~IIID.
The nonspherically and spherically decaying components of the emission
are compared in Section~IV, and a summary is given in Section~V.

\section{Preamble: formulation of the problem}
The physical principles underlying the (mostly unexpected) consequences of the
processes described in Section~I are more transparent if we begin with a
descriptive account of the building blocks of the (unavoidably lengthy) analysis
that follows in Section~III.  Consider a {\it localized} charge $q$ (a charge with linear dimensions much smaller than the
typical radiation wavelengths considered) which moves on a circle of
radius $r$ with the constant angular velocity $\omega{\hat{\bf e}}_z$,
i.e.\ whose path ${\bf x}(t)$ is given, in terms of the cylindrical polar coordinates $(r,\varphi,z)$, by
$$r={\rm const.},\quad z={\rm const.}, \quad \varphi={\hat\varphi}+\omega t,\eqno(1)$$
where ${\hat{\bf e}}_z$ is the basis vector associated with $z$, and ${\hat\varphi}$
the initial value of $\varphi$.

Having defined the path of what will constitute the basic element of our volume source,
we shall use the remainder of this section to describe (A)~the
associated Li\'enard-Wiechert field, (B)~the bifurcation surface which divides the volume of the source into parts with differeing influences on the field, (C)~the Hadamard regularization technique
for dealing with the singularities which are inevitably encountered in the fields
from a superluminal source, (D)~the focal regions in which the nonspherically decaying
part of the emission is detectable and (E)~the relationship between the polarization of the source and of the emission.
\subsection{The Li\'enard-Wiechert fields}
The Li\' enard-Wiechert electric and magnetic fields that arise from a superluminally moving charge $q$ with the trajectory described in Eq.\ (1) are given by
$${\bf E}({\bf x}_P,t_P)=q\sum_{t_{\rm ret}}\bigg[{(1-\vert{\dot{\bf x}}\vert^2/c^2)({\hat{\bf n}}-{\dot{\bf x}}/c)\over\vert1-{\hat{\bf n}}\cdot{\dot{\bf x}}/c\vert^3R^2(t)}+{{\hat{\bf n}}{\bf\times}\{({\hat{\bf n}}-{\dot{\bf x}}/c){\bf\times}{\ddot{\bf x}}\}\over c^2\vert1-{\hat{\bf n}}\cdot{\dot{\bf x}}/c\vert^3R(t)}\bigg]\eqno(2)$$
and ${\bf B}={\hat{\bf n}}{\bf\times}{\bf E}$.
Here, the coordinates $({\bf x}_P,t_P)=(r_P,\varphi_P,z_P,t_P)$ mark
the space-time of observation points, ${\dot{\bf x}}\equiv d{\bf x}/dt$, $c$
is the speed of light {\it in vacuo}, and $R$ and ${\hat{\bf n}}$ are the magnitude
$$R(t)=[(z_P-z)^2+{r_P}^2+r^2-2r_Pr\cos(\varphi_P-{\hat\varphi}-\omega t)]^{1\over2}\eqno(3)$$
and the direction ${\hat{\bf n}}\equiv{\bf R}/R$ of the vector ${\bf R}(t)\equiv{\bf x}_P-{\bf x}(t)$.
The summation extends over all values of the retarded time, i.e.\ all
solutions $t_{\rm ret}<t_P$ of $h(t)\equiv t+R(t)/c=t_P$.
This summation and the absolute-value signs in the factor $\vert1-{\hat{\bf n}}\cdot{\dot{\bf x}}/c\vert$,
which stems from the evaluation of the Dirac delta function $\delta(t-t_P+R/c)$ in the
classical expression for the retarded potential [Eq.\ (10) below], are omitted in most
textbook derivations of Li\'enard-Wiechert fields because in the subluminal regime
the retarded time is a single-valued function of the observation time
and $1-{\hat{\bf n}}\cdot{\dot{\bf x}}/c$ is everywhere positive~\cite{HA4}.

For a given point source $(r, {\hat\varphi}, z)$ with $r\omega>c$ and various
positions $(r_P, \varphi_P, z_P)$ of the observation point, the dependence
$t_P=h(t)$ of the reception time $t_P$ on the emission time $t$ can have one of the generic
forms shown in Fig.\ 1.
\begin{figure}[tbp]
   \centering
\includegraphics[height=7cm]{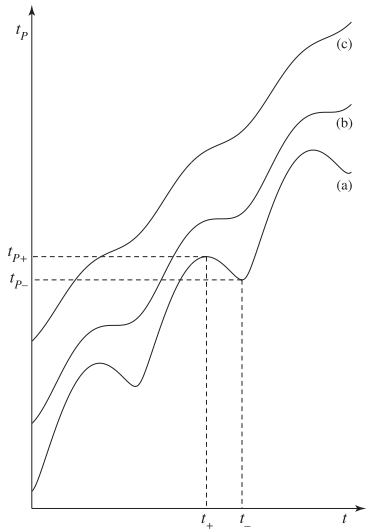}
\caption{The relationship between the observation time $t_P$ and
the emission time $t$ for an observation point that lies (a) inside or on,
(b) on the cusp of, and (c) outside the envelope of the wave fronts or the
bifurcation surface shown in Figs.\ 2 and 3.
This relationship is given by $t_P=t+R(t)/c\equiv h(t;r,\varphi,z; r_P,\varphi_P,z_P)$,
an equation that applies to the envelope when the position $(r,\varphi,z)$ of the source
point is fixed and to the bifurcation surface when the location $(r_P,\varphi_P,z_P)$
of the observer is fixed.  The maxima and minima of curve (a), at which $dR/dt=-c$,
occur on the sheets $\phi_+$ and $\phi_-$ of the envelope or the bifurcation surface,
respectively (Figs.\ 2 and 3).  The inflection points of curve (b), at which $d^2R/dt^2=0$, occur on the
cusp curve of the envelope or the bifurcation surface (Fig.\ 4).}
\end{figure}
As can be seen from curve (a) of this figure, there are values $t_\pm$ of the retarded time at which
$$1-{\hat{\bf n}}\cdot{\dot{\bf x}}/c={\dot h}(t)=1-r_P(r\omega/c)\sin(\varphi_P-\varphi+\omega t)/R(t)=0,\eqno(4)$$ 
i.e.\ at which the source approaches the observer with the speed of light along the radiation
direction ${\hat{\bf n}}$.  The set of observation points at which the field receives
contributions from the values $t_\pm$ of the retarded time are those located on the envelope
of the wave fronts emanating from the moving source in question (Fig.\ 2).
In the vicinity of the extrema of curve (a) in Fig.\ 1, the transcendental equation $t_P=h(t)$
for the retarded time reduces to
$$t_P=t_{P\pm}+\textstyle{1\over2}{\ddot h}(t_\pm)(t-t_\pm)^2+\cdots,\eqno(5)$$where $t_{P\pm}\equiv h(t_\pm)$ are the values of $t_P$ at which the waves emitted at $t=t_\pm$ arrive, and constructively interfere, at the envelope of the wave fronts (see Appendix C of~\cite{HA3}).\par
Were it to exist, a superluminally rotating {\it point} source would therefore generate an
infinitely large field on the envelope of the wave fronts that would emanate from it:
the Li\' enard-Wiechert fields diverge at the extrema of curve (a) in Fig.\ 1, where the
factor $1-{\hat{\bf n}}\cdot{\dot{\bf x}}/c$ in the denominator of Eq.\ (2) vanishes.
However, superluminal sources are necessarily extended and so it is only a superposition
of the Li\' enard-Wiechert fields of their individual volume elements that is physically meaningful.
The relevant quantity is the integral of the above field over the volume of $(r,{\hat\varphi},z)$ space that
the moving localized source occupies in its rest frame.

\begin{figure}[tbp]
   \centering
\includegraphics[height=7cm]{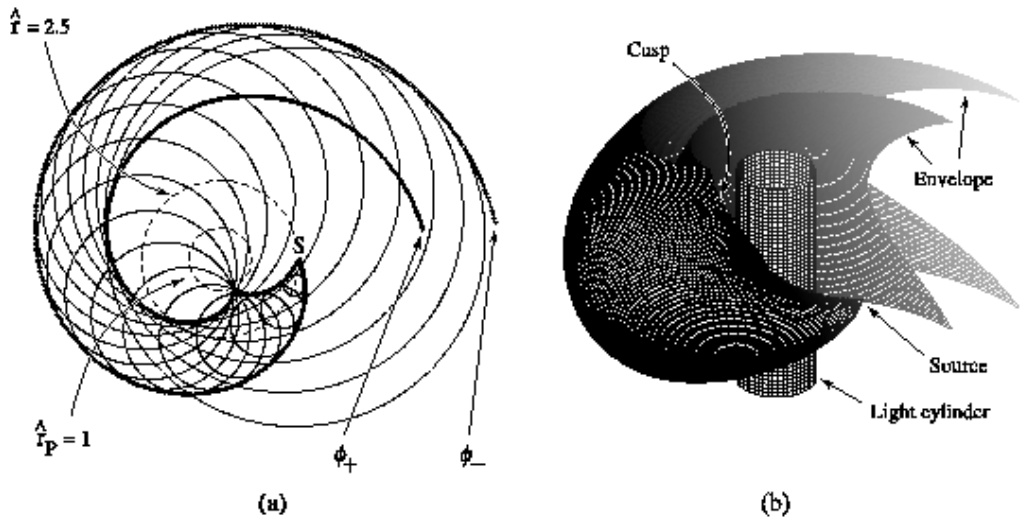}
\caption{(a) Envelope of the spherical wave fronts emanating from a source point $S$ which
moves with a constant angular velocity $\omega$ on a circle of radius
$r=2.5 c/\omega$ (${\hat r}\equiv r\omega/c=2.5$).
The circles in broken lines designate the orbit of $S$ and the light cylinder
$r_P=c/\omega$ (${\hat r}_P=1$).  The curves to which the emitted wave
fronts are tangent are the cross sections of the two sheets $\phi_\pm$ of the envelope
with the plane of source's orbit.  (b) Three-dimensional view of the light cylinder
and the envelope of wave fronts for the same source point $S$.
The tube-like surface constituting the envelope is symmetric with respect to the plane of the orbit.
The cusp along which the two sheets of this envelope meet touches, and is tangential
to, the light cylinder at a point on the plane of the source's orbit and spirals
around the rotation axis out into the radiation zone.  It approaches the
cone $\theta_P=\arcsin(1/{\hat r})$ as $R_P$ tends to infinity
($R_P$, $\theta_P$ and $\varphi_P$ are the spherical coordinates of the observation point $P$).}
\end{figure}

\begin{figure}[tbp]
   \centering
\includegraphics[height=7cm]{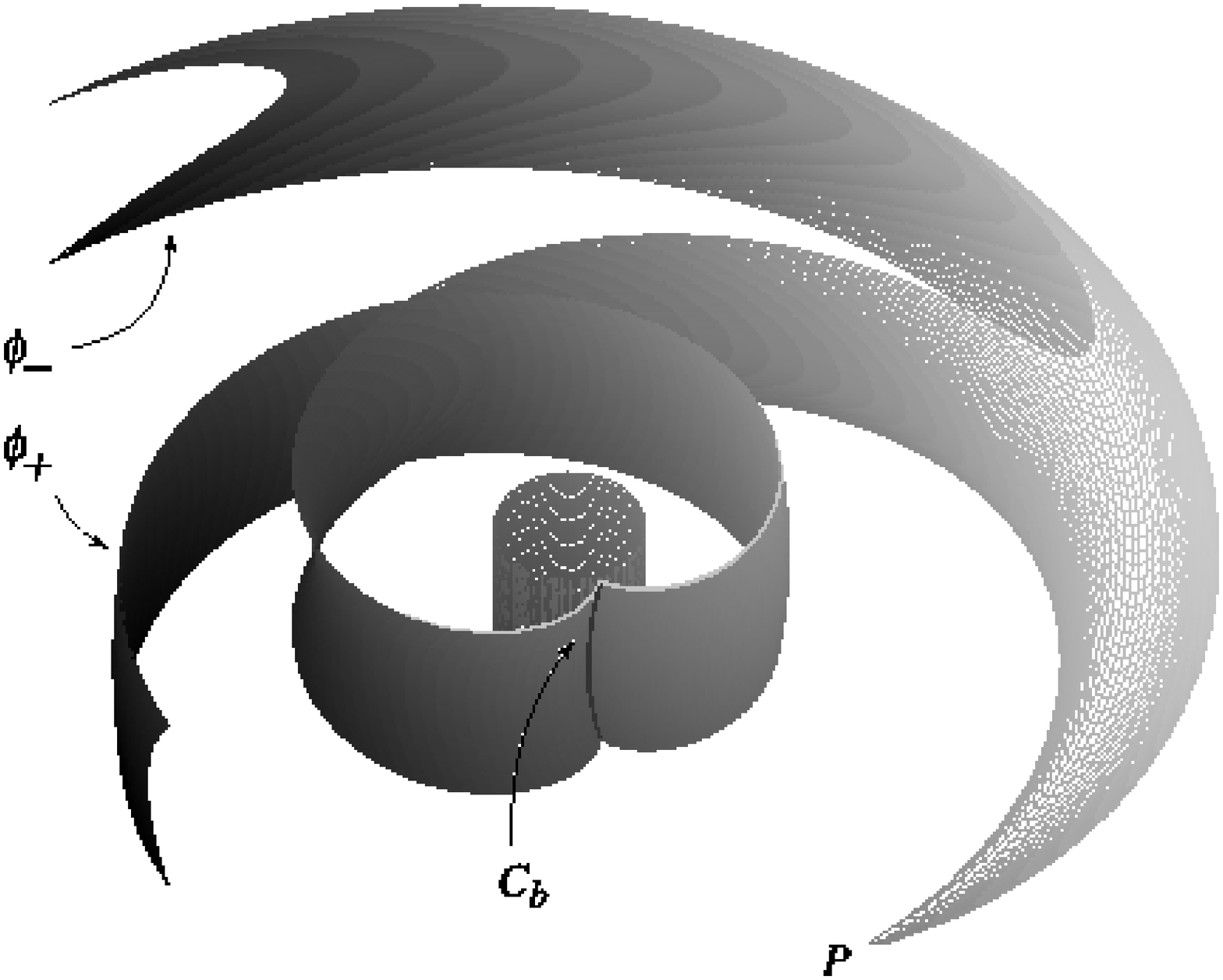}
\caption{The bifurcation surface (i.e.\ the locus of source points that approach the
observer along the radiation direction with the speed of light at the retarded time)
associated with the observation point $P$ at the observation time $t_P$
(the motion of the source is clockwise).  The cusp $C_b$, along which the two
sheets of the bifurcation surface meet, touches and is tangent to the light cylinder
(${\hat r}=1$) at a point on the plane passing through $P$ normal to the rotation axis.
This cusp curve is the locus of source points which approach the observer not only with the speed of light
($dR/dt=-c$) but also with zero acceleration ($d^2R/dt^2=0$) along the radiation direction.
For an observation point in the radiation zone, the spiralling surface that issues from
$P$ undergoes a large number of turns, in which its two sheets intersect one
another, before reaching the light cylinder.}
\end{figure}
\subsection{The bifurcation surface of an observation point}
Considering the field of a given point source, we have so far kept $(r,{\hat\varphi},z)$
fixed and have varied $(r_P,{\hat\varphi}_P,z_P)$.
If we now keep the observation point fixed and allow the coordinates $(r,{\hat\varphi},z)$
of the moving source point to sweep the rest-frame volume of an extended source,
then the curves in Fig.\ 1 would represent the forms the relationship $t_P=h(t)$ assumes
in different regions of the $(r,{\hat\varphi},z)$ space.
For any given $(r_P,{\hat\varphi}_P,z_P)$, there is a set of source elements of a
volume source which approach the observer along the radiation direction ${\hat{\bf n}}$
with the speed of light at the retarded time, i.e.\ for
which ${\dot h}=1-{\hat{\bf n}}\cdot{\dot{\bf x}}/c$ is zero.
The locus of this set of source points in the $(r,{\hat\varphi},z)$ space is given by the
intersection of the volume of the source with the two-sheeted surface
$t_{P\pm}(r,{\hat\varphi},z;r_P,\varphi_P,z_P)-t_P=0$, a surface that we shall refer to
as the {\it bifurcation surface} of the observation point $P$ (Fig.\ 3).
For all source points in the vicinity of this locus, the relationship between the retarded and the
observation times has the form assumed by curve (a) of Fig.\ 1 in the neighbourhood of its
extrema, i.e.\ is that given in Eq.\ (5).

The two sheets ($\pm$) of the bifurcation surface $t_{P\pm}(r,{\hat\varphi},z;r_P,\varphi_P,z_P)-t_P=0$
meet tangentially along a cusp curve (Figs.\ 3 and 4).
The source points on this cusp curve approach the observer not only with the
wave speed (with $dR/dt=-c$) but also with zero acceleration ($d^2R/dt^2=0$)
along the radiation direction.  For such source points, the relationship between the
retarded and the observation times has the form assumed by curve (b) of Fig.\ 1 in
the neighbourhood of its inflection point.

\begin{figure}[tbp]
   \centering
\includegraphics[height=7cm]{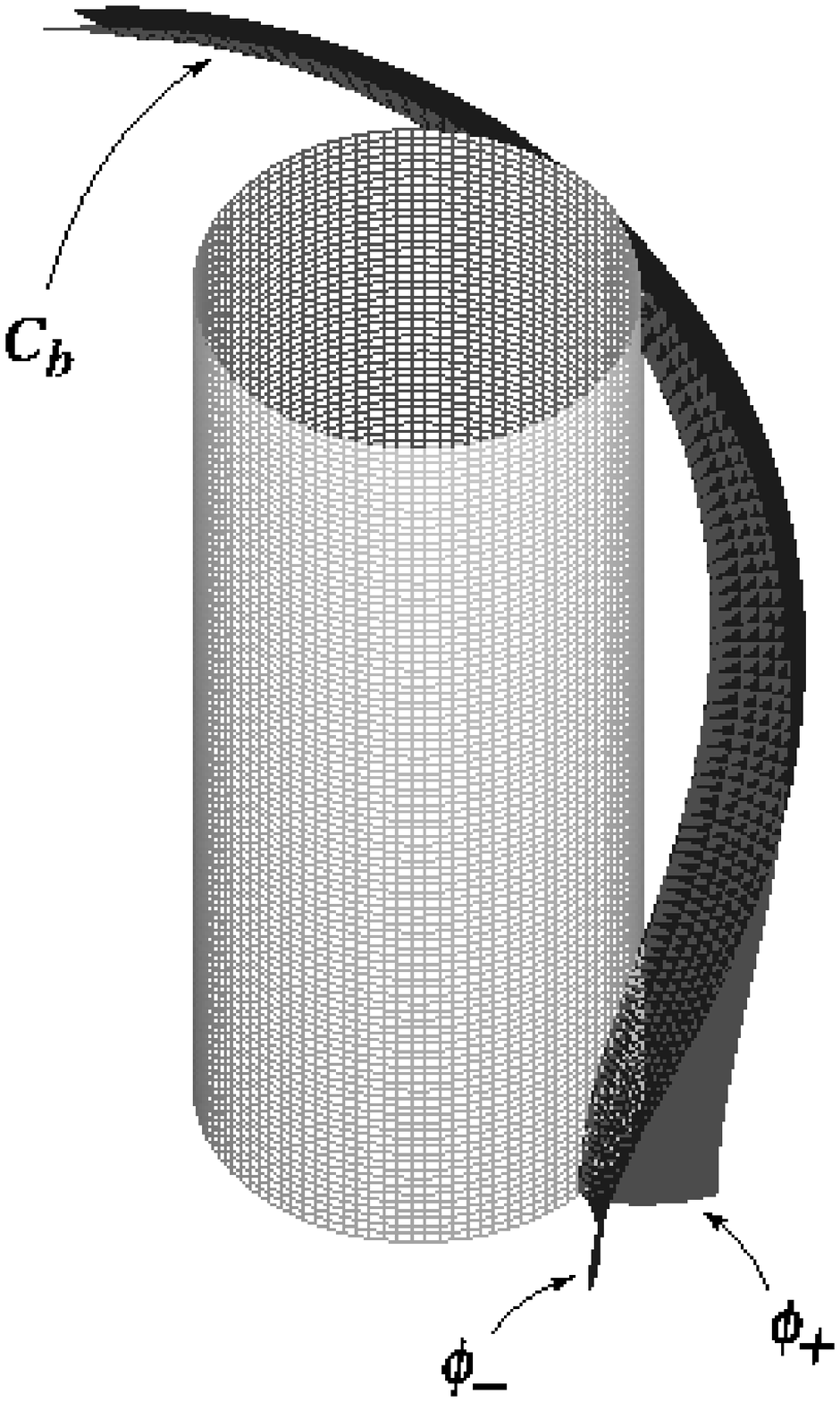}
\caption{Close up of a segment of the cusp curve appearing in Fig.\ 3.
The figure shows the section $0<{\hat z}-{\hat z}_P<5$ of the light cylinder (${\hat r}=1$)
and the two sheets $\phi_\pm$ of the bifurcation surface (the locus of source points that approach the
observer along the radiation direction with the speed of light at the retarded time) in
the vicinity of its cusp curve $C_b$ (the locus of source points that approach the observer
with the speed of light and zero acceleration) for an observer who is located at
${\hat r}_P=3$, ${\hat\varphi}_P=0$.  The cusp curve $C_b$ is symmetrical
with respect to the plane $z=z_P$ passing through the observation point $P$.
The value ${G_j}^{\rm in}$ of the Green's function $G_j$ inside the bifurcation surface
diverges on the inner sides of the two sheets $\phi_+$ and $\phi_-$.
The value ${G_j}^{\rm out}$ of $G_j$ outside the bifurcation surface undergoes
a jump across the strip bordering on the cusp curve onto which these two sheets
coalescence (in the limit ${\hat R}_P\gg1$).}
\end{figure}

The retarded times from which the Li\'enard-Wichert field (2) receives singular contributions
during any given period, therefore, are
$t\simeq t_+\pm[2(t_P+-t_{P+})/{\ddot h}(t_+)]^{1\over2}$ for
 $t_P\,\lsim\,t_{P+}$ and $t\simeq t_-\pm[2(t_P-t_{P-})/{\ddot h}(t_-)]^{1\over2}$
 for $t_P\,\gsim\,t_{P-}$ [see Eq.\ (5)].  Close to these retarded times, the factor
 $1-{\hat{\bf n}}\cdot{\dot{\bf x}}/c$ has the absolute value
$$\vert1-{\hat{\bf n}}\cdot{\dot{\bf x}}/c\vert=\vert{\dot h}(t)\vert\simeq[2{\ddot h}(t_\pm)(t_P-t_{P\pm})]^{1\over2}\eqno(6)$$ 
according to Eq.\ (5).  The functions $t_{P\pm}\equiv t_\pm+R(t_\pm)/c$ in
Eq.\ (6) depend linearly on the coordinate ${\hat\varphi}$ of the source point.
This can be seen from Eq.\ (4) without solving for $t_\pm$:  since $t$ in Eq.\ (4)
appears in only the combination $\varphi_P-{\hat\varphi}-\omega t$,
the solutions $t_\pm$ of this equation are given by expressions of the form
$t_\pm=(\varphi_\pm-{\hat\varphi})/\omega$ in which $\varphi_\pm$ are functions
of $(r,z)$ only.  Given that the functions $R(t_\pm)$ also depend on $t$ through
$\varphi_P-{\hat\varphi}-\omega t$ and so are independent of ${\hat\varphi}$,
it follows from $t_{P\pm}= t_\pm+R(t_\pm)/c$ that $t_{P\pm}-t_P$ has the
functional form $({\hat\varphi}_\pm-{\hat\varphi})/\omega$ in which
${\hat\varphi}_\pm$ depend on $r$ and $z$ only.

According to Eqs.\ (2) and (6), therefore, the Li\'enard-Wiechert fields
diverge like $\vert{\hat\varphi}_\pm-{\hat\varphi}\vert^{-{3\over2}}$
for those source elements close to the bifurcation surface ${\hat\varphi}={\hat\varphi}_\pm$
which approach the observer with the speed of light at the retarded time.
This is a non-integrable singularity.  To superpose the fields of the constituent
volume elements of an extended source we need to integrate the expression
that appears on the right-hand side of Eq.\ (2) over the volume of the $(r,{\hat\varphi},z)$
space occupied by that source.  The multiple integral that needs to be evaluated
here entails an integral with respect to ${\hat\varphi}$ whose integrand is proportional
to $\vert{\hat\varphi}_\pm-{\hat\varphi}\vert^{-{3\over2}}$ in the vicinity of
${\hat\varphi}={\hat\varphi}_\pm$ and so does not exist.
Non-integrable singularities of this type commonly arise in the solutions
of the wave equation over odd-dimensional space-times~\cite{HA5};
they can be handled by the following method from the
theory of generalized functions, a method originally devised by Hadamard~\cite{HA5,HA6,HA6A}.
\subsection{Hadamard's regularization technique}
Hadamard's regularization technique is applicable to situations in which the superposition of the {\it potentials} of the volume elements of an extended source yields a differentiable function of the observer's space-time coordinates, while the superposition of the {\it fields} of those same elements results in a divergent integral.  This technique enables one to extract the physically relevant, finite value of the field of the extended source in question (which would follow from the differentiation of the potential owing to its entire volume) directly from the divergent integral that describes the superposition of the singular fields of its constituent elements.  The calculation and subsequent differentiation of the potential owing to the entire source, a task which can hardly ever be performed analytically in physically realistic situations of this kind, is thus rendered unnecessary.   

In the present case, the singularity of the Li\'enard-Wiechert {\it potential} is like
$\vert1-{\hat{\bf n}}\cdot{\dot{\bf x}}/c\vert^{-1}\sim\vert{\hat\varphi}_\pm-{\hat\varphi}\vert^{-{1\over2}}$
and so is integrable.  The difficulty referred to above
would not arise if we superposed the potentials,
instead of the fields, of the consitituent volume elements of the source.
We would know how to evaluate the required integral over $(r,{\hat\varphi},z)$,
at least in principle, and the function of $({\bf x}_P,t_P)$ that we would thus obtain for the
retarded potential of the entire source could then be differentiated to obtain a finite, singularity-free
expression for the field.  The non-integrable singularity we have encountered stems from
interchanging the orders of integration and differentiation, from differentiating the
Li\' enard-Wiechert potentials of the individual source elements (to obtain the fields)
prior to integrating them over the source volume.  Though feasible in principle, it is not of
course practical to calculate the retarded potential of any physically viable extended
source of the type we are considering explicitly.  The regularization method we are about
to describe is such, however, that the finite value it would assign to the divergent integral
we have encountered exactly equals the value of the field which would follow from directly
differentiating the retarded potential of the extended source~\cite{HA5,HA6}.

All the essential features of the mathematical problem that we face when atempting to
integrate the Li\' enard-Wiechert field (2) with respect to the source coordinate
${\hat\varphi}$ are illustrated by the following simple example.
Consider the function $f(x,a)=(a-x)^{-{1\over2}}$.  If we first evaluate the integral
$I(a)=\int_0^a f(x,a)dx$ of this function with respect to $x$ and then differentiate the
result ($I=2a^{1\over2}$) with respect to $a$ we obtain the well-defined quantity
$dI/da=a^{-{1\over2}}$.  But if we first differentiate $f(x,a)$ with respect to $a$
and then attempt to integrate the resulting function $\partial f/\partial a=-{1\over2}(a-x)^{-{3\over2}}$
over $x$, we obtain
$$J(a)\equiv\int_0^a dx\,\partial f/\partial a=-(a-x)^{-{1\over2}}\big\vert_{x=a}+a^{-{1\over2}},\eqno(7)$$
a quantity which is divergent.  The value which Hadamard's regularization
technique picks out as the finite part of the divergent integral $J(a)$ is $a^{-{1\over2}}=dI/da$.

When applied to the more general form $\int_0^a dx\,F(x)(a-x)^{-{3\over2}}$ of the above divergent
integral, in which $F(x)$ is a regular function, Hadamard's procedure consists of performing an
integration by parts,
\begin{eqnarray*}
\int_0^a dx\,(a-x)^{-{3\over2}}F(x)=&2(a-x)^{-{1\over2}}F(x)\big\vert_a\cr
&-2a^{-{1\over2}}F(0)-2\int_0^a dx\,(a-x)^{-{1\over2}}\partial F(x)/\partial x,\quad&(8)\cr
\end{eqnarray*}
and discarding the term which is divergent.
The remaining finite part [consisting of the last two terms in Eq.\ (8)] is the value which
Hadamard's regularization assigns to this integral; it is the value one would obtain if one first
evaluated $-2\int_0^a dx\,F(x)(a-x)^{-{1\over2}}$ and then differentiated the result with respect to $a$.

The integral arising from the superposition of the Li\'enard-Wiechert fields of the constituent
volume elements of a superluminally rotating charge distribution has an integrand which, in contrast
to that of the integral in Eq.\ (8), is singular {\it within} the domain of integration (rather than on
the boundary of this domain).  Although such cases are seldom encountered in physics, their treatment using Hadamard's regularization is well established in the theory of hyperbolic partial differential equations~\cite{HA5,HA6}.  [In the case of \v Cerenkov emission from an extended source, where there are no contributions from the source elements outside the bifurcation surface (the inverted \v Cerenkov cone issuing from the observation point), the corresponding singularity occurs on the boundary of the domain of integration.]  
Here, we have a divergent integral whose Hadamard finite part consists of two terms,
an integrated term which turns out to decay like ${R_P}^{-{1\over2}}$
with the distance $R_P$ from the source as $R_P$ tends to infinity, and an integral identical to the
classical expression for the retarded field of a volume source [Eq.\ (14) of~\cite{HA7}]
which decays spherically, like ${R_P}^{-1}$.
\subsection{The nonsphericallly decaying part of the emission}
The nonspherically decaying component of the
radiation is detectable only within a limited region of space and during a limited interval of time:
only when the cusps of the envelopes of wave fronts that emanate from the superluminally moving
volume elements of the source propagate past the observer (see Fig.\ 2).

The radiated field entails, at any instant during this limited time interval, a set of wave envelopes
(each associated with the wave fronts emanating from a specific member of a corresponding set of
source elements) whose cusps pass through the position of the privileged observer in question.  These
caustics arise from those volume elements of the source which approach the observer,
along the radiation direction, with the speed of light and zero acceleration at the retarded time
(Figs.\ 3 and 4).  It is the contribution toward the intensity of the field from the filamentary locus
of such source elements that decays like ${R_P}^{-1}$ instead of ${R_P}^{-2}$.

The nonspherical decay of the field does not contravene conservation of energy.  The focused wave
packets that embody the nonspherically decaying pulses are constantly dispersed and reconstructed
out of other waves, so that the constructive interference of their constituent waves takes place
within different solid angles on spheres of different radii $R_P$ (see Appendix D of~\cite{HA3}).
The integral of the flux of energy across a large sphere centred on the source is the same
as the integral of the flux of energy across any other sphere that encloses the source.  The strong
fields that occur in focal regions are compensated by weaker fields elsewhere, so that the
{\it distribution} of the flux of energy across such spheres is highly non-uniform and $R_P$-dependent.
\subsection{Polarization of the source}
This paper is specifically concerned with the spectral and polarization properties of the nonspherically
decaying component of the radiation from a superluminal source.
To assist in identifying the origins of the various polarization components
in the emitted radiation, we base the analysis
on a representative polarization current ${\bf j}=\partial{\bf P}/\partial t$ for which
$$P_{r,\varphi,z}(r, \varphi, z, t)=s_{r,\varphi,z}(r, z)\cos(m{\hat\varphi})\cos(\Omega t),\qquad -\pi<{\hat\varphi}\leq\pi,\eqno(9a)$$
with
$${\hat\varphi}\equiv\varphi-\omega t,\eqno(9b)$$
where $P_{r,\varphi,z}$ are the cylindrical components of the polarization (the electric dipole moment
per unit volume), ${\bf s}(r,z)$ is an arbitrary vector that vanishes outside a finite region of the
$(r, z)$ space and $m$ is a positive integer.
Equation (9) generalizes the earlier
calculation reported in~\cite{HA3}, which was concerned with a rotating charge distribution, to
a case in which the emitting polarization current flows in the $r$ and $z$ as well as in the $\varphi$ direction
and has a distribution pattern that oscillates in addition to moving.
This is similar to the polarization currents available within recently constructed
machines built to study the physics of superluminal emission (see~\cite{AA1,AA2,AA3,AA4,AA5,AA6}
and Appendix A of~\cite{HA7}).

\begin{table}[tbp] \centering
\begin{tabular}{|l|l|}
\hline
 Symbol  & Definition \\ \hline
 $\omega$ &  Angular rotation frequency of the distibution pattern of the source\\ \hline
 $\Omega$ &  The angular frequency with which the source oscillates (in addition to moving)\\ \hline
 $f$ &  Frequency of the radiation generated by the source \\ \hline
 $n=2\pi f/\omega$ &   The harmonic number associated with the radiation frequency\\ \hline
 $m$ &  The number of cycles of the sinusoidal wave train representing the azimuthal\\
 ~ & dependence of the rotating source distribution [see Eq.\ (9)] around the \\
 ~& circumference of a circle centred on, and normal to, the rotation axis \\ \hline
 $\vert\Omega\pm m\omega\vert$ &  The two frequencies at which the nonspherically decaying component \\
 ~ &  of the radiation from the source described in Eq.\ (9) is emitted.  (The spectrum of the \\
 ~ &  corresponding spherically decaying component of the radiation is limited to these two \\
 ~ &  frequencies only if $\Omega/\omega$ is an integer.)\\ \hline
\end{tabular}
\caption{Definition of the various frequencies and numbers
used to describe the source and the emitted radiation}
\label{table1}
\end{table}

Note that the ranges of values of both $\varphi$ and $t$ in Eq.\ (9)
are infinite, as in the case of a rotating point source, but the Lagrangian coordinate
${\hat\varphi}$, which labels each source element by its azimuthal position at
$t=0$, lies in an interval of length $2\pi$~\cite{HA7}.
For a fixed value of $t$, the azimuthal dependence of the density (9) along each circle of radius
$r$ within the source is the same as that of a sinusoidal wave train,
with the wavelength $2\pi r/m$, whose $m$ cycles fit around the circumference of the circle smoothly.
As time elapses, this wave train both propagates around each circle with the velocity
$r\omega$ and oscillates in its amplitude with the frequency $\Omega$ (Table I).
The vector ${\bf s}$ is here left arbitrary in order that we may investigate
the polarization of the resulting radiation for all possible directions of the emitting
current (Table II).  Note that one can construct any distribution with a uniformly
rotating pattern, $P_{r,\varphi,z}(r, {\hat\varphi}, z)$, by the superposition over $m$
of terms of the form $s_{r,\varphi,z}(r, z, m)\cos(m{\hat\varphi})$.

\section{Detailed mathematical treatment of the problem}
\subsection{Integral representation of the Green's function}
In the absence of boundaries, the retarded potential $A^\mu$ arising from any localized distribution of
charges and currents with a density $j^\mu$ is given by
$$A^\mu({\bf x}_P,t_P)=c^{-1}\int d^3x\,dt\,j^\mu({\bf x},t)\delta(t_P-t-R/c)/R,\quad \mu=0,\cdots,3\eqno(10)$$
where $\delta$ is the Dirac delta function, $R$ stands for the magnitude of
${\bf R}\equiv{\bf x}_P-{\bf x}$, and $\mu=1, 2, 3$ designate the spatial components,
${\bf A}$ and ${\bf j}$, of $A^\mu$ and $j^\mu$ in a Cartesian coordinate system.
For the purposes of calculating the electromagnetic fields
$${\bf E}=-{\bf\nabla}_PA^0-\partial{\bf A}/\partial(ct_P)\qquad{\rm and}\qquad {\bf B}={\bf\nabla}_P{\bf\times A}\eqno(11)$$
generated by the source in Eq.\ (9), the space-time of source points may be
marked either with $({\bf x},t)=(r,\varphi,z,t)$ or with the coordinates
$(r,{\hat\varphi},z,t)$ that naturally appear in the description of that rotating source.
In fact, once ${\hat\varphi}$ is adopted as the coordinate that ranges over $(-\pi,\pi)$,
the retarded position $\varphi$ of the rotating source point $(r, {\hat\varphi},z)$
as well as the retarded time $t$ could be used as the coordinate whose range is unlimited.

The electric current density ${\bf j}=\partial{\bf P}/\partial t$ that arises from the polarization
distribution (9) is given, in terms of $\varphi$ and ${\hat\varphi}$, by the real part of
$${\bf j}=\textstyle{1\over2}{\rm i}\omega\sum_{\mu=\mu_\pm}\mu\exp[-{\rm i}(\mu{\hat\varphi}-\Omega\varphi/\omega)]{\bf s},\quad-\pi<{\hat\varphi}\le\pi,\eqno(12)$$
where $\mu_\pm\equiv(\Omega/\omega)\pm m$.
Changing the variables of integration in Eq.\ (10) from $({\bf x}, t)=(r, \varphi, z, t)$ to
$(r, \varphi, z, {\hat\varphi})$ and replacing ${\bf j}$ by the expression in Eq.\ (12), we obtain
$${\bf A}=\textstyle{1\over2}{\rm i}(\omega/c)\sum_{\mu=\mu_\pm}\int_V dV\,\mu\exp(-{\rm i}\mu{\hat\varphi}){\bf s}\int_{\Delta\varphi}d\varphi\,\exp({\rm i}\Omega\varphi/\omega)\delta(g-\phi)/R(\varphi),\eqno(13)$$
where $dV\equiv rdrd{\hat\varphi}dz$.  Here
$\phi$ stands for ${\hat\varphi}-{\hat\varphi}_P$ with ${\hat\varphi}_P\equiv \varphi_P-\omega t_P$,
$R(\varphi)$ is
$$R(\varphi)=[(z_P-z)^2+{r_P}^2+r^2-2r_Pr\cos(\varphi_P-\varphi)]^{1\over2},\eqno(14)$$
the function $g$ is defined by
$$g\equiv\varphi-\varphi_P+{\hat R}(\varphi),\eqno(15)$$
with ${\hat R}\equiv R\omega/c$, $\Delta\varphi$ is the interval of azimuthal
angle traversed by the source, and $V$ is the volume occupied by the source in the
$(r, {\hat\varphi}, z)$ space.

Terms of the order of $R^{-2}$ in ${\bf\nabla}_P{\bf\times A}$, which do not
contribute toward the flux of energy at infinity, may be discarded, as usual~\cite{HA4},
if we are concerned only with the radiation field.
Since the problem we will be considering entails the formation of caustics,
however, we need to treat the arguments of the delta function and its derivative in
the resulting expression for ${\bf B}$ exactly.  Approximating
${\bf\nabla}_P[R^{-1}\delta(t-t_P+R/c)]$ by $R^{-1}\delta^\prime(t-t_P+R/c){\bf\nabla}_P(R/c)$,
i.e.\ discarding the term that arises from the differentiation of $R^{-1}$, we can therefore write
the magnetic field of the radiation as
\begin{eqnarray*}
{\bf B}\simeq&\textstyle{1\over2}{\rm i}(\omega/c)^2\sum_{\mu=\mu_\pm}\int_V dV\,\mu\exp(-{\rm i}\mu{\hat\varphi})\int_{\Delta\varphi}d\varphi\,\exp({\rm i}\Omega\varphi/\omega){\hat{\bf n}}\times{\bf s}\cr
&\times\delta^\prime(g-\phi)/R(\varphi),&(16)\cr
\end{eqnarray*}
where ${\hat{\bf n}}={\bf R}(\varphi)/R(\varphi)$ and
$\delta^\prime$ denotes the derivative of the delta function with respect to its argument.

To put the source density ${\bf s}=s_r{\hat{\bf e}}_r+s_\varphi{\hat{\bf e}}_\varphi+s_z{\hat{\bf e}}_z$
into a form suitable for inserting in Eq.\ (16), we need to express the $\varphi$-dependent
base vectors $({\hat{\bf e}}_r, {\hat{\bf e}}_\varphi, {\hat{\bf e}}_z)$ associated with the source
point $(r, \varphi, z)$ in terms of the constant base vectors
$({\hat{\bf e}}_{r_P},{\hat{\bf e}}_{\varphi_P}, {\hat{\bf e}}_{z_P})$ at the
observation point $(r_P, \varphi_P,z_P)$:
$$\left[\matrix{{\hat{\bf e}}_r\cr {\hat{\bf e}}_\varphi\cr {\hat{\bf e}}_z\cr}\right]=\left[\matrix{\cos(\varphi-\varphi_P)&\sin(\varphi-\varphi_P)&0\cr
-\sin(\varphi-\varphi_P)&\cos(\varphi-\varphi_P)&0\cr
0&0&1\cr}\right]\left[\matrix{{\hat{\bf e}}_{r_P}\cr {\hat{\bf e}}_{\varphi_P}\cr {\hat{\bf e}}_{z_P}\cr}\right].\eqno(17)$$
Equation (17) together with the far-field value of ${\hat{\bf n}}$,
$$\lim_{R\to\infty}{\hat{\bf n}}=\sin\theta_P{\hat{\bf e}}_{r_P}+\cos\theta_P{\hat{\bf e}}_{z_P},\quad\theta_P\equiv\arctan(r_P/z_P),\eqno(18)$$
yields the following expression for the source term in Eq.\ (16):
\begin{eqnarray*}
{\hat{\bf n}}\times{\bf s}=&[s_r\cos\theta_P\cos(\varphi-\varphi_P)-s_\varphi\cos\theta_P\sin(\varphi-\varphi_P)-s_z\sin\theta_P]{\hat{\bf e}}_\parallel\cr
&+[s_\varphi\cos(\varphi-\varphi_P)+s_r\sin(\varphi-\varphi_P)]{\hat{\bf e}}_\perp,&(19)\cr
\end{eqnarray*}
where ${\hat{\bf e}}_\parallel\equiv {\hat{\bf e}}_{\varphi_P}$
(which is parallel to the plane of rotation) and
${\hat{\bf e}}_\perp\equiv{\hat{\bf n}}{\bf\times}{\hat{\bf e}}_\parallel$ comprise
a pair of unit vectors normal to the radiation direction ${\hat{\bf n}}$.

Inserting Eq.\ (19) in Eq.\ (16), rewriting $\delta^\prime(g-\phi)$ as
$-\partial\delta(g-\phi)/\partial{\hat\varphi}$ and making use of the fact that
${\hat{\bf n}}$, ${\bf s}$ and $\Delta\varphi$ are independent of ${\hat\varphi}$, we arrive at
$${\bf B}\simeq-\textstyle{1\over2}{\rm i}(\omega/c)^2\sum_{\mu=\mu_\pm}\int_V dV\,\mu\exp(-{\rm i}\mu{\hat\varphi})\sum_{j=1}^3{\bf u}_j\partial G_j/\partial{\hat\varphi},\eqno(20)$$
where 
$${\bf u}_1\equiv s_r\cos\theta_P{\hat{\bf e}}_\parallel+s_\varphi{\hat{\bf e}}_\perp,\quad{\bf u}_2\equiv -s_\varphi\cos\theta_P{\hat{\bf e}}_\parallel+s_r{\hat{\bf e}}_\perp,\quad{\bf u}_3\equiv-s_z\sin\theta_P{\hat{\bf e}}_\parallel,\eqno(21)$$
and $G_j$ ($j=1,2,3$) are the functions resulting from the remaining integration with respect to $\varphi$:
$$\left[\matrix{G_1\cr G_2\cr G_3\cr}\right]=\int_{\Delta\varphi} d\varphi\,{\delta(g-\phi)\over R}\exp({\rm i}\Omega\varphi/\omega)\left[\matrix{\cos(\varphi-\varphi_P)\cr \sin(\varphi-\varphi_P)\cr 1\cr}\right].\eqno(22)$$
The corresponding expression for the electric field is given by
${\bf E}\simeq{\hat{\bf n}}{\bf \times B}$, as in any other radiation problem.

The functions $G_i(r,{\hat\varphi}, z; r_P,{\hat\varphi}_P, z_P, \varphi_P)$ here act as Green's functions:
they describe the fields of uniformly rotating point sources with fixed (Lagrangian)
coordinates $(r,{\hat\varphi}, z)$ whose strengths sinusoidally vary with time.
In the special case in which $\Omega=0$, i.e.\ the strength of the source is constant,
$G_3$ reduces to the Green's function called $G_0$ in~\cite{HA3} and represents the Li\'enard-Wiechert potential of the point source described in Eq.\ (1).  This may be seen by noting that the evaluation of the delta function in Eq.\ (22) yields
$$G_3\big\vert_{\Omega=0}=\sum_{\varphi=\varphi_j}{1\over R\vert\partial g/\partial\varphi\vert},\eqno(23)$$
where $\varphi_j$ are the solutions of the transcendental equation
$g(\varphi)=\phi$, solutions that are related to those of the equation
$h(t)=t+R(t)/c=t_P$ for the retarded times via $\varphi={\hat\varphi}+\omega t$.
The curves representing $g(\varphi)$ versus $\varphi$ have precisely the same
forms as those appearing in Fig.\ 1.

Similarly, the vector $\partial(G_2\cos\theta_P{\hat{\bf e}}_\perp+G_1{\hat{\bf e}}_\parallel)/\partial{\hat\varphi}$
is proportional to the Li\'enard-Wiechert field (2) when
$\Omega$ is zero:  the electric current associated with the rotating point source
from which field (2) arises flows in the azimuthal direction and so corresponds
to $s_r=s_z=0$.  The singularity structures of $G_i$ are determined by the zeros of
$\partial g/\partial\varphi$ or ${\dot h}$ [see Eq.\ (4)] and so are identical
to the singularity structure already outlined in connection with $G_0$ in~\cite{HA3}.
\subsection{Asymptotic expansion of the Green's function in the time domain}
The retarded times at which the value of the Green's function (23)
[or that of the Li\'enard-Wiechert field (2)] receives divergent contributions from
the point source $(r,{\hat\varphi},z)$ are given by the following solutions of
$\partial g/\partial\varphi=0$ [or, equivalently, of Eq.\ (4)]:
$$\varphi_\pm=\varphi_P+2\pi-\arccos[(1\mp\Delta^{1\over2})/({\hat r}{\hat r}_P)],\eqno(24)$$
with
$$\Delta\equiv({{\hat r}_P}^2-1)({\hat r}^2-1)-({\hat z}-{\hat z}_P)^2\eqno (25)$$
in which $({\hat r}, {\hat z}; {\hat r}_P, {\hat z}_P)$ stand for
($r\omega/c$, $z\omega/c$; $r_P\omega/c$, $z_P\omega/c$).
Note that for a given observation point $(r_P,\varphi_P,z_P,t_P)$,
the critical times $t_\pm=(\varphi_\pm-{\hat\varphi})/\omega$ exist
(i.e.\ are real) only when the $(r,z)$ coordinates of the source point lie within the region
$\Delta\ge0$ of the $(r,z)$ space shown in Fig.\ 5.

\begin{figure}[tbp]
   \centering
\includegraphics[height=7cm]{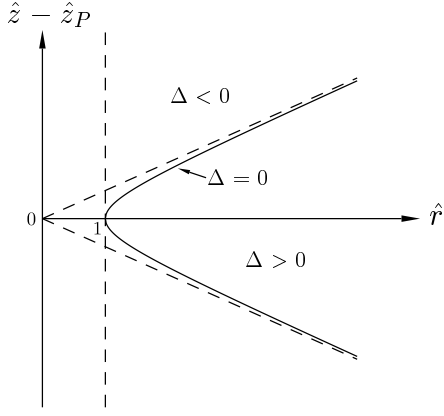}
\caption{The projection $\Delta=0$ of the cusp curve of the bifurcation surface onto
$(r,z)$ space.  The two sheets $\phi_\pm(r,z)$ of the bifurcation surface exist only for
the values of $(r,z)$ in $\Delta\ge0$.  For a given $(r,z)$ in $\Delta\ge0$, the
source point $(r,{\hat\varphi},z)$ lies within the bifurcation surface for
$\phi_-<{\hat\varphi}-{\hat\varphi}_P<\phi_+$ and outside this surface for other values of
${\hat\varphi}$.  The source elements whose $(r,z)$ coordinates fall in
$\Delta<0$ approach the observer with a speed $dR/dt<c$ at the retarded time
and so make contributions toward the field that are no different from those made
in the subluminal regime.  The projection of the cusp curve of the envelope of
wave fronts onto the $(r_P,z_P)$ space is also given by $\Delta=0$ and
has the same shape:  the function $\Delta$ is invariant under the transformation
$(r,z;r_P,z_P)\to(r_P,z_P;r,z)$.  That the cusp curve of the envelope
approaches the cone $\theta_P=\arcsin(1/{\hat r})$ as $R_P$ tends to
infinity can be seen here from the slopes of the asymptotes of the curve $\Delta=0$:
these asymptotes, and so the segments ${\hat z}>{\hat z}_P$ and ${\hat z}<{\hat z}_P$
of the cusp curve, lie in the plane ${\hat z}={\hat z}_P$ for ${\hat r}=1+$
but open up and tend towards the vertical as ${\hat r}$ becomes increasingly
greater than $1$.}
\end{figure}

The locus of source points in the $(r,{\hat\varphi},z)$ space for which
 $\partial g/\partial\varphi=0$ can be found by inserting $t_\pm=(\varphi_\pm-{\hat\varphi})/\omega$
 in the equation $t_P-t-R/c=0$ which specifies the retarded times,
 or equivalently by inserting Eq.\ (24) in $g-\phi=0$ [see Eqs.\ (10), (13) and (15)].  The result is
$$\phi=\phi_\pm\equiv g(\varphi_\pm)=2\pi-\arccos[(1\mp\Delta^{1\over2})/({\hat r}{\hat r}_P)]+{\hat R}_\pm,\eqno(26)$$
where
$${\hat R}_\pm\equiv {\hat R}(\varphi_\pm)=[({\hat z}-{\hat z}_P)^2+{\hat r}^2+{{\hat r}_P}^2-2(1\mp\Delta^{1\over2})]^{1\over2}.\eqno(27)$$ 
The two sheets ($\pm$) of the tube-like spiralling surface
${\hat\varphi}={\hat\varphi}_P+\phi_\pm(r,z)$ meet tangentially and form a cusp where
$\Delta=0$ (Figs.\ 3 and 4).  This cusp curve on which $\partial^2g/\partial\varphi^2$
(and ${\ddot h}$) as well as $\partial g/\partial\varphi$ (and ${\dot h}$) are zero,
constitutes the locus of source points which approach the observer, not only with the speed of
light, but also with zero acceleration along the radiation direction.
On it, the coalescence of two neighbouring stationary points of the phase function
[curve (a) of Fig.\ 1] results in a point of inflection [curve (b) of Fig.\ 1] and so in a
stronger singularity of the Green's function.

For source points in the vicinity of this cusp curve, a unifrom asymptotic approximation to
the value of the integral (22) defining the $G_i$ can be found by the method of
Chester, Friedman and Ursell~\cite{HA8,HA8A}.  Where it is analytic
(i.e.\ for all ${\bf x}\neq{\bf x}_P$), the function $g(\varphi)$ in the argument of
the delta function in Eq.\ (22) may be transformed into the following cubic function:
$$g(\varphi)=\textstyle{1\over3}\nu^3-{c_1}^2\nu+c_2,\eqno(28)$$
where $\nu$ is a new variable of integration replacing $\varphi$,
and the coefficients $c_1$ and $c_2$ are chosen such that the values of the two
functions on opposite sides of Eq.\ (28) coincide at their extrema:
$$c_1\equiv(\textstyle{3\over4})^{1\over3}(\phi_+-\phi_-)^{1\over3},\quad c_2\equiv\textstyle{1\over2}(\phi_++\phi_-).\eqno(29)$$
Insertion of Eq.\ (28) in Eq.\ (22) results in
$$G_j=\int_{\Delta\nu}d\nu\, f_j(\nu)\delta(\textstyle{1\over3}\nu^3-{c_1}^2\nu+c_2-\phi)\eqno(30)$$
where
$$\left[\matrix{f_1\cr f_2\cr f_3\cr}\right]=R^{-1}(d\varphi/d\nu)\exp({\rm i}\Omega\varphi/\omega)\left[\matrix{\cos(\varphi-\varphi_P)\cr \sin(\varphi-\varphi_P)\cr 1\cr}\right],\eqno(31)$$
and $\Delta\nu$ is the image of $\Delta\varphi$ under transformation (28).\par
The leading term in the asymptotic expansion of the integral (30) for small
$c_1$ can now be obtained~\cite{HA8,HA9,HA10,HA10A} by replacing its integrand $f_j$ with $p_j+q_j\nu$
and extending its range $\Delta\nu$ to $(-\infty,\infty)$:
$$G_j\sim\int_{-\infty}^\infty d\nu\, (p_j+q_j\nu)\delta(\textstyle{1\over3}\nu^3-{c_1}^2\nu+c_2-\phi),\eqno(32)$$
where
$$p_j=\textstyle{1\over2}(f_j\vert_{\nu=c_1}+f_j\vert_{\nu=-c_1}),\eqno(33a)$$
$$q_j=\textstyle{1\over2}{c_1}^{-1}(f_j\vert_{\nu=c_1}-f_j\vert_{\nu=-c_1}),\eqno(33b)$$
and the symbol $\sim$ denotes asymptotic approximation.
(Note that the critical points $\varphi=\varphi_\pm$ transform into $\nu=\mp c_1$, respectively.)
This integral has precisely the same form as that evaluated in Appendix A of~\cite{HA3}.

It behaves differently inside ($\phi_-<\phi<\phi_+$) and outside ($\phi<\phi_-,\, \phi>\phi_+$)
the bifurcation surface so that
$$G_j=\cases{{G_j}^{\rm in}&$\vert\chi\vert<1$\cr
           {G_j}^{\rm out}&$\vert\chi\vert>1$,\cr}\eqno(34a)$$
with
$${G_j}^{\rm in}\sim2{c_1}^{-2}(1-\chi^2)^{-{1\over2}}[p_j\cos(\textstyle{1\over3}\arcsin\chi)-c_1q_j\sin(\textstyle{2\over3}\arcsin\chi)],\eqno(34b)$$
$${G_j}^{\rm out}\sim{c_1}^{-2}(\chi^2-1)^{-{1\over2}}[p_j\sinh(\textstyle{1\over3}{\rm arccosh}\vert\chi\vert)+c_1q_j{\rm sgn}(\chi)\sinh(\textstyle{2\over3}{\rm arccosh}\vert\chi\vert)],\eqno(34c)$$
and $\chi\equiv3(\phi-c_2)/(2{c_1}^3)$ [cf.\ Eqs.\ (A16) and (A17) of~\cite{HA3}].
[The two-dimensional loci $\chi=\pm1$ across which each $Gj$ changes
form correspond, according to Eq.\ (29), to the two sheets $\phi_\pm$ of
the bifurcation surface, respectively.]  Explicit expressions for the coefficients
$p_j(r,z)$ and $q_j(r,z)$ are given in Appendix A.

The above results show that as a source point $(r,{\hat\varphi}, z)$ in the vicinity of
the cusp curve approaches the bifurcation surface from inside,
i.e.\ as $\chi\to1-$ or $\chi\to-1+$, ${G_j}^{\rm in}$ and hence $G_j$ diverges.
However, as a source point approaches one of the sheets of the bifurcation
surface from outside, $G_j$ tends to a finite limit:
$${G_j}^{\rm out}\big\vert_{\phi=\phi_\pm}={G_j}^{\rm out}\big\vert_{\chi=\pm1}\sim(p_j\pm2c_1q_j)/(3{c_1}^2),\eqno(35)$$
for the numerator of ${G_j}^{\rm out}$ is also zero when $\vert\chi\vert=1$.
The Green's function $G_j$ is singular, in other words, only on the
inner side of the bifurcation surface (see Fig.\ 6).

\begin{figure}[tbp]
   \centering
\includegraphics[height=7cm]{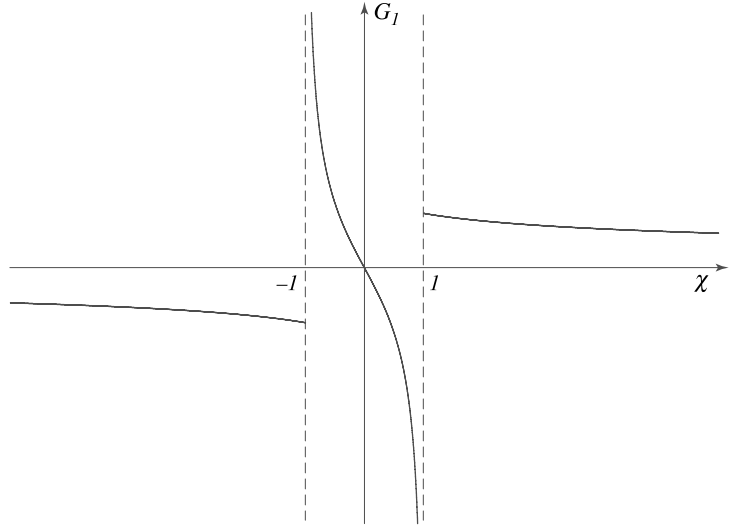}
\caption{The $\chi$ dependence of the Green's function $G_1$ in the far-field limit where $p_1=0$.
The values of this function inside and outside the interval $-1<\chi<1$
represent ${G_1}^{\rm in}$ and ${G_1}^{\rm out}$, respectively:
the space inside the bifurcation surface [the interval $\phi_-<\phi<\phi_+$ at a given
$(r,z)$] is mapped into $\vert\chi\vert<1$ and that outside the bifurcation
surface into $\vert\chi\vert>1$.  The function ${G_1}^{\rm in}$ diverges on
the inner sides, $\chi=1-$ and $\chi=-1+$, of the two sheets $\phi_+$ and $\phi_-$
of the bifurcation surface.  The function ${G_1}^{\rm out}$ is discontinuous
across a two-dimensional strip bordering on the cusp curve of the bifurcation surface:
even at points close to this cusp curve where the separation $\phi_+-\phi_-$
of the two sheets of the bifurcation surface tends to zero (Figs.\ 3 and 4),
the vanishingly small interval $\phi_-<\phi<\phi_+$ in $\phi$ is mapped into the
finite interval $-1<\chi<1$ in $\chi$, so that the difference
${G_1}^{\rm out}\vert_{\chi=1+}-{G_1}^{\rm out}\vert_{\chi=-1-}$
between the values of the function ${G_1}^{\rm out}$ on opposite sides of the
strip in question remains non-zero.}
\end{figure}

Moreover, like the diffraction field near a focal point~\cite{HA9}, the
Green's function ${G_j}^{\rm out}$ undergoes a phase shift across the coalescent
surfaces $\phi=\phi_\pm$ at the cusp curve (Fig.\ 4 and 6).
The shift in the sign of the second term in Eq.\ (34c) results in a finite discontinuity in the value of
${G_j}^{\rm out}$ across the strip bordering on the cusp curve where the
two sheets of the bifurcation surface are tangential:  even in the limit
$c_1\to 0$, where $\phi_-$ and $\phi_+$ are coincident,
${G_j}^{\rm out}\vert_{\phi=\phi_+}-{G_i}^{\rm out}\vert_{\phi=\phi_-}$
has the non-vanishing value ${4\over3}q_j/c_1$.  It is this discontinuity
in the value of the Green's function (the {\it potential} of a point source)
across the cusp curve of the bifurcation surface that gives rise to nonspherically
decaying boundary contributions to the {\it field} of a volume source.
(Note that a discontinuity of this type, which resembles that of a step function,
cannot be handled by means of an analysis in the frequency domain
unless the contributions from an infinitely large number of Fourier components
of the decomposed Green's function are accurately superposed again when evaluating the field.)
\subsection{Hadamard's finite part of the integral representing the radiation field}
The integral which represents the superposition of the contributions of the source
elements in the vicinity of the cusp curve of the bifurcation surface to the value of
the magnetic field, i.e.\ the integral resulting from the insertion of Eq.\ (34) in Eq.\ (20),
is divergent:  the derivative $\partial G_j/\partial{\hat\varphi}$ of the Green's
function has a singularity [$\sim(1-\chi^2)^{-{3\over2}}$] that is not integrable
(with respect to $\chi$ or equivalently ${\hat\varphi}$).

Our task in this section is to extract the Hadamard finite part ~\cite{HA5,HA6}
of this divergent integral, the finite quantity that we would have obtained had we
been able to evaluate the potential ${\bf A}({\bf x}_P,t_P)$ of the entire source
explicitly prior to applying the operator ${\bf\nabla}_P\times$ to this potential.

The function $\partial G_j/\partial{\hat\varphi}$ appearing under the integral sign in
Eq.\ (20) is given by different expressions in different regions of
$(r,{\hat\varphi},z)$ space [see Eq.\ (34)].  For an observation point, the cusp
curve of whose bifurcation surface intersects the source distribution, therefore,
the domain of integration in Eq.\ (20) has to be divided into a part $V_{\rm in}$
that lies within and a part $V_{\rm out}$ that lies without the bifurcation surface
before the integrand of the integral in this equation can be written out explicitly (Fig.\ 3).
The Green's function $G_j$ has no singularities in the region $\Delta<0$ of the $(r,z)$
space (Fig.\ 5), so that the contribution of the source elements in the part of $V_{\rm out}$
for which $\Delta<0$ are no different from the contributions of the source elements
that lie in the subluminally moving portion of the source [see Eqs.\ (23)--(25)].
It would be sufficient to consider the contributions only of those source elements
for which $\Delta\ge0$.  All source elements for which $\partial G_j/\partial{\hat\varphi}$
is singular are taken into account once the integration with respect to
${\hat\varphi}$ (at fixed values of $r$ and $z$ in $\Delta\ge0$)
is performed over the following two intervals:
the interval $\phi_-<{\hat\varphi}-{\hat\varphi}_P<\phi_+$ in which
$\vert\chi\vert<1$ and $G_j={G_j}^{\rm in}$,
and the remaining part of $-\pi<{\hat\varphi}<\pi$ in which
$\vert\chi\vert>1$ and $G_j={G_j}^{\rm out}$ (see Fig.\ 6).

Thus the magnetic field ${\bf B}_{\Delta\ge0}$ that arises from the
source elements in $\Delta\ge0$ can be written, according to Eq.\ (20), as ${\bf B}^{\rm in}+{\bf B}^{\rm out}$ with
$${\bf B}^{\rm in,out}=-\textstyle{1\over2}{\rm i}(\omega/c)^2\sum_{j=1}^3\int_{\Delta\ge0}r\,dr\,dz\,{\bf u}_j{K_j}^{\rm in,out},\eqno(36a)$$
where
$${K_j}^{\rm in}\equiv\sum_{\mu=\mu_\pm}\int_{\phi_-}^{\phi_+}d\phi\,\mu\exp(-{\rm i}\mu{\hat\varphi})\partial {G_j}^{\rm in}/\partial{\hat\varphi},\eqno(36b)$$
and
$${K_j}^{\rm out}\equiv\sum_{\mu=\mu_\pm}\Big(\int_{-\pi-{\hat\varphi}_P}^{\phi_-}+\int_{\phi_+}^{\pi-{\hat\varphi}_P}\Big)d\phi\,\mu\exp(-{\rm i}\mu{\hat\varphi})\partial {G_j}^{\rm out}/\partial{\hat\varphi}.\eqno(36c)$$
(Recall that $dV\equiv rdrd{\hat\varphi}dz$ and $\phi\equiv{\hat\varphi}-{\hat\varphi}_P$.)\par
Once it is integrated by parts, the integral in Eq.\ (36b) in turn splits into two terms:
$${K_j}^{\rm in}=\sum_{\mu=\mu_\pm}\Big\{\Big[\mu\exp(-{\rm i}\mu{\hat\varphi}){G_j}^{\rm in}\Big]_{\phi_-}^{\phi_+}+{\rm i}\int_{\phi_-}^{\phi_+}d\phi\,{\mu}^2\exp(-{\rm i}\mu{\hat\varphi}){G_j}^{\rm in}\Big\},\eqno(37)$$
of which the first (integrated) term is divergent [see Eq.\ (34) and Fig.\ 6].
Hadamard's finite part of ${K_j}^{\rm in}$ and hence of ${\bf B}^{\rm in}$ (here designated by
the prefix ${\cal F}$) is obtained by discarding this divergent contribution toward the value of
${K_j}^{\rm in}$ (see~\cite{HA5,HA6,HA6A}):
$${\cal F}\big\{{K_j}^{\rm in}\big\}={\rm i}\sum_{\mu=\mu_\pm}\int_{\phi_-}^{\phi_+}d\phi\,{\mu}^2\exp(-{\rm i}\mu{\hat\varphi}){G_j}^{\rm in}.\eqno(38)$$
Note that the singularity of the kernel of this integral, i.e.\ the singularity
of ${G_j}^{\rm in}$, is like that of $\vert{\hat\varphi}_\pm-{\hat\varphi}\vert^{-{1\over2}}$ and so is integrable.

The boundary contributions that result from the integration of the right-hand side of
Eq.\ (36c) by parts are well-defined automatically:
\begin{eqnarray*}
{K_j}^{\rm out}=&\sum_{\mu=\mu_\pm}\Big\{\Big[\mu\exp(-{\rm i}\mu{\hat\varphi}){G_j}^{\rm out}\Big]_{\phi_-}^{\phi_+}\cr
&+{\rm i}\Big(\int_{-\pi-{\hat\varphi}_P}^{\phi_-}+\int_{\phi_+}^{\pi-{\hat\varphi}_P}\Big)d\phi{\mu}^2\exp(-{\rm i}\mu{\hat\varphi}){G_j}^{\rm out}\Big\},&(39)\cr
\end{eqnarray*}
for ${G_j}^{\rm out}$ tends to a finite limit as the bifurcation surface is approached from
outside (Fig.\ 6) and ${G_j}^{\rm out}\vert_{\phi=\pi-{\hat\varphi}_P}$ equals
${G_j}^{\rm out}\vert_{\phi=-\pi-{\hat\varphi}_P}$ when $\phi_\pm\ne\pm\pi-{\hat\varphi}_P$.
The integral representing ${K_j}^{\rm out}$, in other words, is finite by itself and needs no regularization.

If we now insert ${\cal F}\{{K_j}^{\rm in}\}$ and ${K_j}^{\rm out}$ from Eqs.\ (38) and
(39) in Eq.\ (36a) and combine ${\bf B}^{\rm in}$ and ${\bf B}^{\rm out}$, we arrive at an
expression for the Hadamard finite part of ${\bf B}_{\Delta\ge0}$ which entails both a
volume and a surface integral:  ${\cal F}\{{\bf B}_{\Delta\ge0}\}={\bf B}^{\rm s}+{\bf B}^{\rm ns}$.
The volume integral
$${\bf B}^{\rm s}=\textstyle{1\over2}(\omega/c)^2\sum_{\mu=\mu_\pm}\mu^2\int_{\Delta\ge0}r\,dr\,dz\,\int_{-\pi}^\pi d{\hat\varphi}\,\exp(-{\rm i}\mu{\hat\varphi})\sum_{j=i}^3{\bf u}_jG_j\eqno(40)$$
has the same form as the familiar integral representation of the field
of a subluminal source~\cite{HA7} and decays spherically
(like ${R_P}^{-1}$ for $R_P\to\infty$).  The surface integral
$${\bf B}^{\rm ns}\equiv-\textstyle{1\over2}{\rm i}(\omega/c)^2\sum_{j=1}^3\int_{\Delta\ge0}r\,dr\,dz\,{\bf u}_j{K_j}^{\rm boundary}\eqno(41)$$
stems from the boundary contribution
$${K_j}^{\rm boundary}\equiv\sum_{\mu=\mu_\pm}\Big[\mu\exp(-{\rm i}\mu{\hat\varphi}){G_j}^{\rm out}\Big]_{\phi_-}^{\phi_+}\eqno(42)$$
in Eq.\ (39) and, as we shall see below, turns out to decay nonspherically
(like ${R_P}^{-{1\over2}}$ for $R_P\to\infty$).

Making use of Eq.\ (35) to rewrite ${G_j}^{\rm out}\vert_{\phi_-}^{\phi_+}$
in Eq.\ (42) explicitly, we obtain
\begin{eqnarray*}
{K_j}^{\rm boundary}=&\textstyle{2\over3}\sum_{\mu=\mu_\pm}\mu\exp[-{\rm i}\mu({\hat\varphi}_P+c_2)]\big[2{c_1}^{-1}q_j\cos(\textstyle{2\over3}\mu{c_1}^3)\cr
&-{\rm i}{c_1}^{-2}p_j\sin(\textstyle{2\over3}\mu{c_1}^3)\big],&(43)\cr
\end{eqnarray*}
where $c_1$ and $c_2$ are defined in Eq.\ (29).  The asymptotic expansion of
${G_j}^{\rm out}$ in Eq.\ (34c) is for small $c_1$.  To be consistent, therefore,
we must likewise replace the above expression by the first
term of its Taylor expansion in powers of $c_1$:
$${K_j}^{\rm boundary}=\textstyle{4\over3}{c_1}^{-1}q_j\sum_{\mu=\mu_\pm}\mu\exp[-{\rm i}\mu({\hat\varphi}_P+\phi_-)]+\cdots,\eqno(44)$$
for the remainder of this series is by a factor of the order of ${c_1}^2$
smaller than the above retained term.

The value of $c_1$ close to the cusp curve of the bifurcation surface
(where $\Delta=0$) is in turn given by
$$c_1=2^{-{1\over3}}({\hat r}^2{{\hat r}_P}^2-1)^{-{1\over2}}\Delta^{1\over2}+O(\Delta).\eqno(45)$$
This may be obtained by using Eq.\ (25) to express ${\hat z}$
everywhere in Eqs.\ (26) and (27) in terms of $\Delta$ and ${\hat r}$ and
expanding the resulting expressions in powers of $\Delta^{1\over2}$.
When the observation point lies in the far zone, ${\hat r}$ on the cusp curve
of the bifurcation surface has the value $\csc\theta_P$ [see Eq.\ (25)] and so $c_1$
assumes the value $2^{-{1\over3}}{{\hat R}_P}^{-1}\Delta^{1\over2}$.
Moreover, according to Eqs.\ (A9) and (A10),
$$q_j\simeq2^{2\over3}(\omega/c){{\hat R}_P}^{-1}\exp[{\rm i}(\Omega/\omega)(\varphi_P+3\pi/2)]{\bar q}_j\eqno(46a)$$
with
$${\bar q}_j\equiv\big(1\quad-{\rm i}\Omega/\omega\quad{\rm i}\Omega/\omega\big)\eqno(46b)$$
for $\Delta\ll1$ and ${\hat R}_P\gg1$ (see Appendix A).

Equations (41), (44), (45) and (46) thus jointly yield
\begin{eqnarray*}
{\bf B}^{\rm ns}\simeq&-\textstyle{4\over3}{\rm i}\exp[{\rm i}(\Omega/\omega)(\varphi_P+3\pi/2)]\sum_{\mu=\mu_\pm}\mu\exp(-{\rm i}\mu{\hat\varphi}_P)\cr
&\times\sum_{j=1}^3{\bar q}_j\int_{\Delta\ge0}{\hat r}\,d{\hat r}\,d{\hat z}\,\Delta^{-{1\over2}}{\bf u}_j\exp(-{\rm i}\mu\phi_-).&(47)\cr
\end{eqnarray*}
The integrand of the surface integral in Eq.\ (47) is still singular on the projection
of the cusp curve of the bifurcation surface onto the $(r,z)$ space (Fig.\ 5)
but this singularity is integrable.  
\subsection{Amplitude, frequencies and polarization of the nonspherically decaying radiation outside the plane of source's orbit}
The leading term in the asymptotic expansion of the
integral in Eq.\ (47) for high $\mu_\pm$ (i.e.\ for the regime in which the wavelengths
of oscillations of the source are much shorter than the length scale $c/\omega$ of its orbit)
may be obtained by the method of stationary phase provided that $\theta_P\ne\pi/2$.

Both derivatives, $\partial\phi_-/\partial {\hat r}$ and $\partial\phi_-/\partial {\hat z}$,
of the function that appears in the phase of the rapidly oscillating exponential in
Eq.\ (47) vanish at the point ${\hat r}=1$, ${\hat z}={\hat z}_P$, where
the cusp curve of the bifurcation surface touches, and is tangential to, the light cylinder (see Figs.\ 3 and 4).
However, $\partial^2\phi_-/\partial{\hat r}^2$ diverges at this point, so that
neither the phase nor the amplitude of the integrand of the integral in
Eq.\ (47) are analytic at ${\hat r}=1,{\hat z}={\hat z}_P$.
Only for an observer who is located outside the plane of rotation, i.e.\ whose
coordinate $z_P$ does not match the coordinate $z$ of any source element,
is the function $\phi_-$ analytic throughout the domain of integration.
To take advantage of the simplifications offered by the analyticity of
$\phi_-$ as a function of ${\hat r}$, we proceed under the assumption that $\theta_P\ne\pi/2$.

Since $\phi_-\equiv g(\varphi_-)$ and $\partial g/\partial\varphi=\partial g/\partial r=0$ along the curve 
\begin{eqnarray*}
C:\quad&{\hat r}={\hat r}_C({\hat z})\equiv\{\textstyle{1\over2}({{\hat r}_P}^2+1)-[\textstyle{1\over4}({{\hat r}_P}^2-1)^2-({\hat z}-{\hat z}_P)^2]^{1\over2}\}^{1\over2},\cr
&\varphi=\varphi_C({\hat z})\equiv\varphi_P+2\pi-\arccos({\hat r}_C/{\hat r}_P)&(48)\cr
\end{eqnarray*}
[Eqs.\ (15) and (26)], $\phi_-$ is stationary as a function of
${\hat r}$, i.e.\  $\partial\phi_-/\partial{\hat r}=0$, on the projection
${\hat r}_C({\hat z})$ of $C$ onto the $(r,z)$ plane (see also~\cite{HA7}).
In the far-field limit, where the terms $({\hat z}-{\hat z}_P)^2/({{\hat r}_P}^2-1)^2$
and ${\hat r}_C/{\hat r}_P$ in Eq.\ (48) are much smaller than unity, curve $C$ coincides with the locus
\begin{eqnarray*}
C_b:\quad&{\hat r}=[1+({\hat z}-{\hat z}_P)^2/({{\hat r}_P}^2-1)]^{1\over2},\cr
&\varphi=\varphi_P+2\pi-\arccos[1/({\hat r}{\hat r}_P)],&(49)\cr
\end{eqnarray*}
of source points which approach the observer along the radiation direction with the wave
speed and zero acceleration at the retarded time, i.e.\ coincides with the cusp curve of the
bifurcation surface (Figs.\ 3 and 4).

It can be seen from the far-field limit of Eq.\ (49) that the cusp curve
$C_b$ would intersect the source distribution if $\theta_P$ lies in the interval
$\vert\theta_P-{\pi\over2}\vert\leq\arccos\, (1/{{\hat r}_>})$,
where ${\hat r}_>\equiv r_>\omega/c$ and $r_>$ is the radial coordinate
of the outer boundary of the source~\cite{HA3}.
Hence, the range of values of $\theta_P$ for which the stationary points of
$\phi_-$ fall within the source distribution would be as wide as $(0,\pi)$
if the extent $r_>-c/\omega$ of the superlumially moving part of the source is
comparable to the radius $c/\omega$ of the light cylinder (see Fig.\ 5).

The dominant terms in the Taylor expansion of $\phi_-$ about a point
$({\hat r}_C, \varphi_C, {\hat z})$ on curve $C$ (with an
arbitrary coordinate ${\hat z}\neq{\hat z}_P$) are
$$\phi_-=\phi_C-[\textstyle{1\over2}({{\hat r}_P}^2-1)({{\hat r}_C}^2-1)^{-1}-1]{{\hat R}_C}^{-1}({\hat r}-{\hat r}_C)^2+\cdots,\eqno(50)$$
in which we have denoted the values of $\phi$ and
${\hat R}$ on $C$ by $\phi_C\equiv{\hat R}_C+\varphi_C-\varphi_P$ and
$${\hat R}_C\equiv[({\hat z}_P-{\hat z})^2+{{\hat r}_P}^2-{{\hat r}_C}^2]^{1\over2},\eqno(51)$$
respectively.  Moreover, $\Delta^{1\over2}$ has the finite value
${{\hat r}_C}^2-1$ on $C$ [Eqs.\ (25) and (48)].

We are now in a position to evaluate the leading term in the asymptotic expansion
of the ${\hat r}$ integral in Eq.\ (47) by applying the principle of stationary phase~\cite{HA10,HA10A}:
by replacing the phase of the rapidly oscillating exponential that appears in this integral
with its Taylor expansion (50) and approximating the amplitude of the
integrand with its value along $C$.  The resulting integral, at a given value of ${\hat z}$, is
\begin{eqnarray*}
\int_{\Delta\ge0}{\hat r}\,d{\hat r}\,&\Delta^{-{1\over2}}{\bf u}_j\exp(-{\rm i}\mu\phi_-)\sim\exp(-{\rm i}\mu\phi_C){\hat r}_C({{\hat r}_C}^2-1)^{-1}{\bf u}_j\big\vert_C\cr
&\times\int_0^{{\hat r}_>-{\hat r}_C} d\eta\,\exp\{{\rm i}\mu[\textstyle{1\over2}({{\hat r}_P}^2-1)({{\hat r}_C}^2-1)^{-1}-1]{{\hat R}_C}^{-1}\eta^2\},&(52)
\end{eqnarray*}
in which ${\hat r}_C$, $\phi_C$ and ${\hat R}_C$ have their far-field values
$${\hat r}_C\simeq\csc\theta_P,\quad\phi_C\simeq3\pi/2+{\hat R}_C,\quad{\hat R}_C\simeq{\hat R}_P-{\hat z}\cos\theta_P,\eqno(53)$$
and $\eta\equiv{\hat r}-{\hat r}_C$ [see Eqs.\ (48) and (51)].
This integral entails an integrand whose phase is large on account of both
$\mu_\pm\gg1$ and ${\hat R}_P\gg1$.  It can be cast into the form of, and
evaluated as, a Fresnel integral with a large argument to arrive at
\begin{eqnarray*}
\int_{\Delta\ge0}{\hat r}\,d{\hat r}\,\Delta^{-{1\over2}}{\bf u}_j\exp(-{\rm i}\mu\phi_-)\sim&(\pi/2)^{1\over2}{{\hat R}_P}^{-{1\over2}}\vert\sin\theta_P\cos\theta_P\vert^{-1}\vert\mu\vert^{-{1\over2}}\cr
&\times\exp[-{\rm i}(\mu\phi_C-\textstyle{\pi\over4}{\rm sgn}\,\mu)]{\bf u}_j\big\vert_C&(54)\cr
\end{eqnarray*}
when $\theta_P\ne\pi/2$.  Note that the values ${\bf u}_j\vert_C$ of ${\bf u}_j$
along the curve $C$ are functions of $z$ and that this curve becomes parallel to the
$z$ axis as $R_P$ tends to infinity (Figs.\ 4 and 5).

Once the surface integration in Eq.\ (47) is expressed in terms
of a double integral, this result may be used to obtain
\begin{eqnarray*}
{\bf B}^{\rm ns}\sim&-\textstyle{4\over3}{\rm i}(2\pi)^{1\over2}{{\hat R}_P}^{-{1\over2}}\vert\sin2\theta_P\vert^{-1}\exp({\rm i}\Omega\varphi_C/\omega)\sum_{\mu=\mu_\pm}\vert\mu\vert^{1\over2}{\rm sgn}(\mu)\exp({\rm i}\textstyle{\pi\over4}{\rm sgn}\,\mu)\cr
&\times\exp[-{\rm i}\mu({\hat R}_P-\omega t_P+\varphi_C)]\sum_{j=1}^3{\bar q}_j\int_{-\infty}^{\infty}d{\hat z}\,{\bf u}_j\big\vert_C\exp({\rm i}\mu{\hat z}\cos\theta_P),&(55)\cr
\end{eqnarray*}
in which $\varphi_C\equiv\varphi_P+3\pi/2$.
The remaining ${\hat z}$ integration in this expression amounts to a
Fourier decomposition of the source densities $s_{r,\varphi,z}\vert_C$.

Insertion of Eqs.\ (21) and (46b) in Eq.\ (55) and the introduction of the Fourier transforms
$${\bar s}_{r,\varphi,z}\equiv\int_{-\infty}^\infty d{\hat z}\, s_{r,\varphi,z}\big\vert_C\exp({\rm i}\mu{\hat z}\cos\theta_P).\eqno(56)$$
result in 
\begin{eqnarray*}
{\bf E}^{\rm ns}\sim&\textstyle{4\over3}(2\pi)^{1\over2}{{\hat R}_P}^{-{1\over2}}\vert\sin2\theta_P\vert^{-1}\exp({\rm i}\Omega\varphi_C/\omega)\sum_{\mu=\mu_\pm}\vert\mu\vert^{1\over2}{\rm sgn}(\mu)\exp({\rm i}\textstyle{\pi\over4}{\rm sgn}\,\mu)\cr
&\times\exp[-{\rm i}\mu({\hat R}_P-\omega t_P+\varphi_C)]\big\{({\rm i}{\bar s}_\varphi+\Omega{\bar s}_r/\omega){\hat{\bf e}}_\parallel\cr
&-[({\rm i}{\bar s}_r-\Omega{\bar s}_\varphi/\omega)\cos\theta_P+\Omega{\bar s}_z\sin\theta_P/\omega]{\hat{\bf e}}_\perp\big\}&(57)\cr
\end{eqnarray*}
for the electric field (${\bf E}^{\rm ns}\sim {\hat{\bf n}}{\bf\times}{\bf B}^{\rm ns}$)
of the nonspherically decaying part of the radiation.
This expression is valid, of course, only for an observation point, outside the plane of rotation, the
cusp curve of whose bifurcation surface intersects the source distribution,
i.e.\ for $0<\vert\theta_P-{\pi\over2}\vert\leq\arccos\, (1/{{\hat r}_>})$.

Thus the spectrum of the nonspherically decaying part of the radiation only contains the frequencies
$\mu_\pm\omega=\Omega\pm m\omega$, i.e.\ the frequencies that enter the creation
or practical implementation of the source described in Eq.\ (9).  This contrasts with the spectrum of the spherically
spreading part of the radiation which extends to frequencies of the order of
$\Omega^3/\omega^2$ when $\Omega/\omega\gg1$ (cf.\ ~\cite{HA7}).
The intensity of the radiation at the frequency $\Omega+m\omega$ is
the same as its intensity at $\Omega-m\omega$ when either $\Omega/\omega\gg m$
or $\Omega/\omega\ll m$.  The radiation at one of these frequencies would have a much
lower intensity, on the other hand, if $\Omega$ and $m\omega$ are comparable in magnitude.

Equation (57) shows, moreover, that the nonspherically decaying component of the
radiation is linearly polarized both for $\Omega/\omega\ll1$ and for
$\Omega/\omega\gg1$ when one of the cylindrical
components of ${\bar{\bf s}}$ is appreciably larger than the others.
In the case of an ${\bf s}$ which lies in the azimuthal direction, the radiation is
polarized parallel to the plane of rotation for $\Omega/\omega\ll1$ and normal to that direction
(and phase shifted by $\pi/2$) for $\Omega/\omega\gg1$.
The plane of polarization of the radiation coincides with the plane passing
through the observer and the rotation axis if ${\bf s}$ lies parallel to the rotation axis.
For this radiation to be elliptically polarized, on the other hand, ${\bar{\bf s}}$
needs to have two cylindrical components that are comparable in magnitude
and $\Omega/\omega$ has to be of the order of unity (Table~\ref{table2}).

\begin{table}[tbp] \centering
\begin{tabular}{|l|l|l|l|}
\hline
 ~ &  ~ & ~ & \\
 ~ & $\frac{\Omega}{\omega} \gg 1$ & $\frac{\Omega}{\omega} \sim 1$ & $\frac{\Omega}{\omega} \ll 1$ \\
 ~ &  ~ & ~ & \\ \hline
$s_r\neq0,s_\varphi=s_z=0$ & linear, ${\hat{\bf e}}_\parallel$, phase$=0$ & elliptic & linear, ${\hat{\bf e}}_\perp$, phase$=-\frac{\pi}{2}$ \\ \hline
$s_\varphi\neq0,s_r=s_z=0$ & linear, ${\hat{\bf e}}_\perp$, phase$=0$ & elliptic & linear, ${\hat{\bf e}}_\parallel$, phase$=+\frac{\pi}{2}$ \\ \hline
$s_z\neq0,s_r=s_\varphi=0$ & linear, ${\hat{\bf e}}_\perp$, phase$=\pi$ &  linear, ${\hat{\bf e}}_\perp$, phase$=\pi$ & linear, ${\hat{\bf e}}_\perp$, phase$=\pi$ \\ \hline
\end{tabular}
\caption{The state of polarization of the nonspherically decaying
component of the emitted radiation for different ranges of $\Omega/\omega$ and different
orientations of the emitting polarization current}
\label{table2}
\end{table}

\section{Discussion: comparison of the nonspherically and spherically decaying components of the radiation}
The radiation field that arises from the superluminal portion ($r>c/\omega$)
of the volume source described in Eq.\ (9) consists, as shown in the preceding section,
of two components:  a nonspherically decaying component whose intensity diminishes like
${R_P}^{-1}$ with the distance $R_P$ from the source and a spherically spreading component,
one whose intensity has the conventional dependence ${R_P}^{-2}$ on $R_P$.
The former is described by Eq.\ (57) and the latter, which follows from Eq.\ (40),
was earlier calculated in~\cite{HA7}.  The nonspherically decaying part of the radiation
${\bf E}^{\rm ns}$ is only emitted at the two frequencies $\mu_\pm\omega$,
whereas the spherically decaying part ${\bf E}^{\rm s}$ has a discrete spectrum,
comprising multiples $n\omega$ of the rotation frequency, which extends as far as
$n\sim(\Omega/\omega)^3$ when $\Omega/\omega$ ($\gg1$) is different from an integer (see Table I).

As in the case of any other linear system, the present emission process generates
an output only at those frequencies which are carried both by its input (the source)
and its response (Green's) function.  The agent responsible for the generation of a
broadband radiation from the source described in Eq.\ (9), whose creation or practical
implementation only entails the two frequencies $\mu_\pm\omega$, is acceleration:
a remarkable effect of centripetal acceleration is to enrich the spectral content of a rotating
volume source, for which $\Omega/\omega$ is different from an integer, by effectively
endowing the distribution of its density with space-time discontinuities.
(For a detailed discussion of this point, see~\cite{HA7}.)

The electric field of the spherically spreading part of the radiation for incommensurate
values of $\Omega$ and $\omega$ is given by the real part of
$${\bf E}^{\rm s}={{\tilde{\bf E}}^{\rm s}}_0+2\sum_{n=1}^{\infty}{{\tilde{\bf E}}^{\rm s}}_n\exp(-{\rm i}n{\hat\varphi}_P), \eqno(58)$$  
in which the Fourier component ${{\tilde{\bf E}}^{\rm s}}_n$ of this field
at the frequency $n\omega$ has the following value beyond the Fresnel zone:
\begin{eqnarray*}
{{\tilde{\bf E}}^{\rm s}}_n\sim&\textstyle{1\over2}{{\hat r}_P}^{-1}\exp\{-{\rm i}[n({\hat R}_P+\textstyle{3\over2}\pi)-(\Omega/\omega)(\varphi_P+\textstyle{3\over2}\pi)]\}({\hat r}_>-{\hat r}_<)Q_{\hat\varphi}{\bar{\bf Q}}_z\cr
&+\{m\to-m,\Omega\to -\Omega\},&(59)\cr
\end{eqnarray*}
with
$$Q_{\hat\varphi}=-\sum_{\mu=\mu_\pm}\mu^2\sin[\pi(n-\mu)]/(n-\mu),\eqno(60)$$
and
\begin{eqnarray*}
{\bar{\bf Q}}_z\equiv&\big[{\bar s}_r{\bf J}_{n-\Omega/\omega}(n)+{\rm i}{\bar s}_\varphi{{\bf J}^\prime}_{n-\Omega/\omega}(n)\big]{\hat{\bf e}}_\parallel+\big[({\bar s}_\varphi\cos\theta_P\cr
&-{\bar s}_z\sin\theta_P){\bf J}_{n-\Omega/\omega}(n)-{\rm i}{\bar s}_r\cos\theta_P{{\bf J}^\prime}_{n-\Omega/\omega}(n)\big]{\hat{\bf e}}_\perp&(61)
\end{eqnarray*}
[cf.\ Eq.\ (66) of~\cite{HA7}].
Here, ${\hat r}_<<1$ and ${\hat r}_>>1$ denote the lower and upper limits of the radial
interval in which the source densities $s_{r,\varphi,z}$ are non-zero, the symbol
$\{m\to-m,\Omega\to-\Omega\}$ designates a term like the one preceding it in
which $m$ and $\Omega$ are everywhere replaced by $-m$ and $-\Omega$,
respectively, and ${\bf J}_{n-\Omega/\omega}(n)$ and
${{\bf J}^\prime}_{n-\Omega/\omega}(n)$ are the Anger function~\cite{HA11}
and the derivative of the Anger function with respect to its argument.

To compare the amplitudes of ${\bf E}^{\rm ns}$ and ${\bf E}^{\rm s}$ at a frequency
with which both these components are emitted, let us consider a case in which
$\Omega/\omega$ equals an integer, so that the spherically decaying part of the
radiation is also emitted only at the frequencies $\mu_\pm\omega$.  In this case,
the quantity $Q_{\hat\varphi}$ in Eq.\ (60) is non-zero only if $n$
equals $\mu_+$ or $\mu_-$, and $Q_{\hat\varphi}\vert_{n=\mu_\pm}=-\pi{\mu_\pm}^2$.

At the higher of the two frequencies, i.e.\ at $\mu_+\omega=\Omega+m\omega$,
the amplitude of ${\bf E}^{\rm ns}$ has the value
\begin{eqnarray*}
\vert{{\bf E}^{\rm ns}}_{\mu_+}\vert\sim&\textstyle{4\over3}(2\pi)^{1\over2}{{\hat R}_P}^{-{1\over2}}\vert\sin2\theta_P\vert^{-1}{\mu_+}^{1\over2}\big\vert({\rm i}{\bar s}_\varphi+\Omega{\bar s}_r/\omega){\hat{\bf e}}_\parallel\cr
&-[({\rm i}{\bar s}_r-\Omega{\bar s}_\varphi/\omega)\cos\theta_P+\Omega{\bar s}_z\sin\theta_P/\omega]{\hat{\bf e}}_\perp\big\vert&(62)\cr
\end{eqnarray*}
according to Eq.\ (57).  This should be compared with the following amplitude implied by Eqs.\ (58)--(61): 
\begin{eqnarray*}
\vert{{\tilde{\bf E}}^{\rm s}}_{\mu_+}\vert\sim&\textstyle{1\over2}\pi{{\hat r}_P}^{-1}({\hat r}_>-{\hat r}_<){\mu_+}^2\big\vert[{\bar s}_rJ_m(\mu_+)+{\rm i}{\bar s}_\varphi{J^\prime}_m(\mu_+)]{\hat{\bf e}}_\parallel\cr
&+[({\bar s}_\varphi\cos\theta_P-{\bar s}_z\sin\theta_P)J_m(\mu_+)-{\rm i}{\bar s}_r\cos\theta_P{J^\prime}_m(\mu_+)]{\hat{\bf e}}_\perp\big\vert&(63)
\end{eqnarray*}
since the Anger functions
${\bf J}_{n-\Omega/\omega}(n)$ and ${{\bf J}^\prime}_{n-\Omega/\omega}(n)$
in Eq.\ (61) respectively reduce to the Bessel functions
$J_m(\mu_+)$ and ${J^\prime}_m(\mu_+)$ when $\Omega/\omega$ is
an interger and so $n$ is exactly equal to $\mu_+$ (see~\cite{HA7,HA11}).

In a case where the emitting polarization current is parallel to the rotation axis,
for instance, the ratio of the amplitudes of the two components of the radiation is given by
$$\vert{{\bf E}^{\rm ns}}_{\mu_+}\vert/\vert{{\tilde{\bf E}}^{\rm s}}_{\mu_+}\vert\sim{\textstyle{4\over3}}(\textstyle{2\over\pi})^{1\over2}\vert\sec\theta_P\vert{\mu_+}^{-{3\over2}}(\Omega/\omega)\vert J_m(\mu_+)\vert^{-1}{{\hat R}_P}^{1\over2}({\hat r}_>-{\hat r}_<)^{-1},\eqno(64)$$
since $s_r$ and $s_\varphi$ would then be zero.
For $\Omega/\omega\gg1$, the amplitude of $J_m(\mu_+)$ is of the
order of ${\mu_+}^{-{1\over2}}$.  Irrespective of whether $m$ is smaller
than or comparable to $\Omega/\omega$, therefore, the factor
${\mu_+}^{-{3\over2}}(\Omega/\omega)\vert J_m(\mu_+)\vert^{-1}$
is independent of frequency.  Thus the above ratio is already much greater than unity
($\sim{\mu_+}^{1\over2}$) at the Fresnel distance $R_P\sim(\mu_+\omega/c)(r_>-r_<)^2$
from the source, a result which holds true, as can be seen from Eqs.\ (57) and (59),
even when $\Omega/\omega$ is different from an integer and $s_r$ and $s_\varphi$ are non-zero.
\section{Summary}
We have examined the electromagnetic emission from
polarization charge-currents whose distribution patterns
have the time dependence of a travelling wave with an accelerated
superluminal motion.
Such macroscopic polarization currents are not incompatible with the requirements
of special relativity because their superluminally moving distribution
patterns are created by the coordinated motion of aggregates of subluminally moving particles.
Our analysis is based on an emitting polarization current,
which has a poloidal as well as a
toroidal component, and which has a distribution pattern that
sinusoidally oscillates in addition to superluminally rotating (Table~I);
similar currents are employed in recently-constructed machines
designed to test the physics of superluminal emission.

We find that such sources
generate localized electromagnetic waves that do not decay
spherically, i.e.\ that do not have an intensity diminishing like ${R_P}^{-2}$
with the distance $R_P$ from their source [Eq.\ (57)].
The nonspherical decay of the focused wave packets that are
emitted does not contravene conservation of energy (Section~IID):
the constructive interference of the constituent waves of such propagating
caustics takes place within different solid angles on spheres of different radii ($R_P$)
centred on the source.

Detailed analysis in the far-field limit shows that the spectrum of the
nonspherically decaying part of the radiation emitted by the source described in Eq.\ (9) only contains the frequencies
$\Omega\pm m\omega$, i.e.\ the frequencies that enter the creation
or practical implementation of that source (Section~IIID and Table~I).  This
contrasts with the spectrum of the spherically
spreading part of the radiation which extends to higher frequencies (Section~IV).

We have also determined the polarization of
the nonspherically decaying component of the
radiation in the far-field limit.
In many cases, the emission is highly linearly polarized [Eq.\ (57) and Table~II];
however, with certain source frequencies and polarizations it is possible to
also produce elliptical polarized emission.

Finally, we have examined the relative amplitudes of the spherically and nonspherically decaying
components of the emission.  We find that even at relatively short distances from the emitter,
the latter component can represent the greater part of the observed signal.
\section{Acknowledgements}
This work is supported by EPSRC grant GR/M52205.
We should like to thank J.M. Rodenburg, D. Lynden-Bell,
W. Hayes and  G.S. Boebinger for helpful comments
and enthusiastic support.
\section*{Appendix: evaluation of the leading terms in the asymptotic expansion of the Green's function}
This appendix concerns the evaluation of the coefficients $p_j(r,z)$ and $q_j(r,z)$
in the asymptotic expansion (34) of the Green's function ${G_j}$ for small $c_1$.
These coefficients are defined, by Eqs.\ (31) and (33), in terms of the functions
$\varphi_\pm$, ${\hat R}_\pm$ and $c_1$, which appear in Eqs.\ (24)--(27) and (29),
and the derivative $d\varphi/d\nu$ which is to be calculated from Eq.\ (28).

We have already seen in Section~IIIC that close to the cusp curve of the bifurcation surface
(where $\Delta=0$) the function $c_1(r,z)$ can be approximated as in Eq.\ (45).
In the regime of validity of the asymptotic expansion (34), where $\Delta$ is much smaller than
$({\hat r}^2{{\hat r}_P}^2-1)^{1\over2}$, the functions
$\varphi_\pm$ and ${\hat R}_\pm$ may likewise be expressed in terms of
$({\hat r},\Delta)$ [instead of $({\hat r},{\hat z})$] and expanded in
powers of $\Delta^{1\over2}$ to arrive at
$$\varphi_\pm=\varphi_c\mp({\hat r}^2{{\hat r}_P}^2-1)^{-{1\over2}}\Delta^{1\over2}+O(\Delta),\eqno(A1)$$
$${\hat R}_\pm=({\hat r}^2{{\hat r}_P}^2-1)^{1\over2}\pm({\hat r}^2{{\hat r}_P}^2-1)^{-{1\over2}}\Delta^{1\over2}+O(\Delta),\eqno(A2)$$
where $\varphi_c\equiv\varphi_P+2\pi-\arccos[1/({\hat r}{\hat r}_P)]$.
Recall that $\nu=\pm c_1$ in Eq.\ (33) are the images under the mapping (28) of
$\varphi=\varphi_\mp$, respectively.

The remaining functions $d\varphi/d\nu\vert_{\nu=\pm c_1}$ that appear in the definitions of
$p_j$ and $q_j$ are indeterminate.  Their corresponding values have to be found
by repeated differentiation of Eq.\ (28) with respect to $\nu$:
$$(dg/d\varphi)(d\varphi/d\nu)=\nu^2-{c_1}^2,\eqno(A3)$$
$$(d^2g/d\varphi^2)(d\varphi/d\nu)^2+(dg/d\varphi)(d^2\varphi/d\nu^2)=2\nu,\eqno(A4)$$
etc., and the evaluation of the resulting relations at $\nu=\pm c_1$.
This procedure, which amounts to applying the l'H\^opital's rule, yields
$$d\varphi/d\nu\vert_{\nu=\pm c_1}=(2c_1{\hat R}_{\mp})^{1\over2}/\Delta^{1\over4}\eqno(A5)$$
and
$$d^2\varphi/d\nu^2\vert_{\nu=\pm c_1}=\pm\textstyle{1\over3}\Delta^{-{1\over4}}(2{\hat R}_\mp/c_1)^{1\over2}\big[1-(2^{1\over3}c_1{\hat R}_\mp/\Delta^{1\over2})^{3\over2}(1\pm3\Delta^{1\over2}/{{\hat R}_\mp}^2)\big]\eqno(A6)$$
for the values of the first two derivatives of $\varphi$.
Close to the cusp curve $\Delta=0$, where $c_1$ and ${\hat R}_\pm$ may
be approximated as in Eqs.\ (45) and (A2), these reduce to
$d\varphi/d\nu\vert_{\nu=0}=2^{1\over3}$ and
$d^2\varphi/d\nu^2\vert_{\nu=0}=-2^{-{1\over3}}({\hat r}^2{{\hat r}_P}^2-1)^{-{1\over2}}$,
so that the contribution of $d^2\varphi/d\nu^2\vert_{\nu=0}$ to $q_j$ is negligible in the far field.

Inserting Eqs.\ (A1), (A2), (A5) and (45) in the defining equations (31) and (33),
keeping only the dominant terms in powers of $\Delta^{1\over2}$ and ${{\hat R}_\pm}^{-1}$,
and taking the far-field limit ${\hat R_P}\gg1$ of the resulting expressions, we obtain
$$p_1\simeq2^{1\over3}(\omega/c){{\hat R}_P}^{-2}\exp({\rm i}\Omega\varphi_c/\omega),\eqno(A7)$$
$$p_2\simeq-2^{1\over3}(\omega/c){{\hat R}_P}^{-1}\exp({\rm i}\Omega\varphi_c/\omega),\quad p_3\simeq-p_2,\eqno(A8)$$
and
$$q_1\simeq2^{2\over3}(\omega/c){{\hat R}_P}^{-1}\exp({\rm i}\Omega\varphi_c/\omega),\eqno(A9)$$
$$q_2\simeq-q_3\simeq-{\rm i}(\Omega/\omega)q_1,\eqno(A10)$$
where $\varphi_c$ in these expressions has its far-field value $\varphi_P+3\pi/2$.
Here, we have made use of the fact that the indeterminate ratio $(\Delta^{1\over2}/c_1)_{\Delta=0}$
has the value $2^{1\over3}({\hat r}^2{{\hat r}_P}^2-1)^{1\over2}$, and that as
$R_P$ tends to infinity ${\hat r}$ approaches the value $\csc\theta_P$
along the cusp curve of the bifurcation surface [see Eq.\ (49)].

\end{document}